\documentclass[12 pt, amsfonts, amssymb,color]{article}

\usepackage{amsmath,amssymb,amsfonts,latexsym}
\usepackage{color} 

\usepackage[all]{xy}
\def\R{\mathbb{R}}

\begin{document}
%<<<<<<<<<<< enumeration of eqns section wise>>>>>>>>>>>>>>>>>>>

\renewcommand\theequation{\arabic{section}.\arabic{equation}}
\catcode`@=11 \@addtoreset{equation}{section}
%<<<<<<<<<<<<<<<<<<<<<<<<<<<<<<<<<>>>>>>>>>>>>>>>>>>>>>>>>>>>>>>>>>
\newtheorem{defn}{Definition}[section]
\newtheorem{theorem}{Theorem}[section]
\newtheorem{axiom2}{Example}[section]
\newtheorem{lem}{Lemma}[section]
\newtheorem{prop}{Proposition}[section]
\newtheorem{cor}{Corollary}[section]
\newcommand{\be}{\begin{equation}}
\newcommand{\ee}{\end{equation}}

\def\forms{{\textstyle\bigwedge}}
\newcommand{\equal}{\!\!\!&=&\!\!\!}
\newcommand{\rd}{\partial}
\newcommand{\g}{\hat {\cal G}}
\newcommand{\bo}{\bigodot}
\newcommand{\res}{\mathop{\mbox{\rm res}}}
\newcommand{\diag}{\mathop{\mbox{\rm diag}}}
\newcommand{\Tr}{\mathop{\mbox{\rm Tr}}}
\newcommand{\const}{\mbox{\rm const.}\;}
\newcommand{\cA}{{\cal A}}
\newcommand{\bA}{{\bf A}}
\newcommand{\Abar}{{\bar{A}}}
\newcommand{\cAbar}{{\bar{\cA}}}
\newcommand{\bAbar}{{\bar{\bA}}}
\newcommand{\cB}{{\cal B}}
\newcommand{\bB}{{\bf B}}
\newcommand{\Bbar}{{\bar{B}}}
\newcommand{\cBbar}{{\bar{\cB}}}
\newcommand{\bBbar}{{\bar{\bB}}}
\newcommand{\bC}{{\bf C}}
\newcommand{\cbar}{{\bar{c}}}
\newcommand{\Cbar}{{\bar{C}}}
\newcommand{\Hbar}{{\bar{H}}}
\newcommand{\cL}{\mathcal{L}}
\newcommand{\bL}{{\bf L}}
\newcommand{\Lbar}{{\bar{L}}}
\newcommand{\cLbar}{{\bar{\cL}}}
\newcommand{\bLbar}{{\bar{\bL}}}
\newcommand{\cM}{{\cal M}}
\newcommand{\bM}{{\bf M}}
\newcommand{\Mbar}{{\bar{M}}}
\newcommand{\cMbar}{{\bar{\cM}}}
\newcommand{\bMbar}{{\bar{\bM}}}
\newcommand{\cP}{{\cal P}}
\newcommand{\cQ}{{\cal Q}}
\newcommand{\bU}{{\bf U}}
\newcommand{\bR}{{\bf R}}
\newcommand{\cW}{{\cal W}}
\newcommand{\bW}{{\bf W}}
\newcommand{\bZ}{{\bf Z}}
\newcommand{\Wbar}{{\bar{W}}}
\newcommand{\Xbar}{{\bar{X}}}
\newcommand{\cWbar}{{\bar{\cW}}}
\newcommand{\bWbar}{{\bar{\bW}}}
\newcommand{\abar}{{\bar{a}}}
\newcommand{\nbar}{{\bar{n}}}
\newcommand{\pbar}{{\bar{p}}}
\newcommand{\tbar}{{\bar{t}}}
\newcommand{\ubar}{{\bar{u}}}
\newcommand{\utilde}{\tilde{u}}
\newcommand{\vbar}{{\bar{v}}}
\newcommand{\wbar}{{\bar{w}}}
\newcommand{\phibar}{{\bar{\phi}}}
\newcommand{\Psibar}{{\bar{\Psi}}}
\newcommand{\bLambda}{{\bf \Lambda}}
\newcommand{\bDelta}{{\bf \Delta}}
\newcommand{\p}{\partial}
\newcommand{\om}{{\Omega \cal G}}
\newcommand{\ID}{{\mathbb{D}}}

\def\ep{\epsilon}
\def\de{\delta}

\def\wt{\widetilde}
\def\fracpd#1#2{\frac{\partial #1}{\partial #2}}
\def\pd#1#2{\frac{\partial #1}{\partial #2}}

\def\<#1>{\langle#1\rangle}

 \def\rc#1{\begin{color}{red}#1\end{color}}
  \def\bc#1{\begin{color}{blue}#1\end{color}}

\title{Infinitesimal time reparametrisation and its applications} 
\author{Jos\'e F. Cari\~nena$^1$\footnote{E-mail:jfc@unizar.es}, Eduardo Mart\'{\i}nez$^2$\footnote{E-mail:emf@unizar.es} and Miguel C. Mu\~noz-Lecanda$^3$\footnote{E-mail:miguel.carlos.munoz@upc.edu}\\
$^1$Departamento de F\'{\i}sica Te\'orica and IUMA, Universidad de Zaragoza,\\Pedro Cerbuna 12, 
E-50009 Zaragoza, Spain\\$^2$Departamento de    Matem\'atica Aplicada and 
IUMA, Universidad de Zaragoza,\\
Pedro Cerbuna 12, 
E-50009 Zaragoza, Spain\\
$^3$Departament de Matem\'atiques, Campus Nord U.P.C., Ed. C-3\\
C/ Jordi Girona 1. E-08034 Barcelona, Spain}

\date{ }

\maketitle

%----------------
%\begin{quote}
%{\tt  [Filename: \jobname.tex] }
%\end{quote}
%----------------

\begin{abstract} 
A geometric approach to Sundman infinitesimal time-reparametrisation is given and some of its applications are used to
 illustrate the general theory. Special emphasis is put on geodesic motions and systems described by mechanical type Lagrangians.
 The Jacobi metric  appears as a particular case of a  Sundman transformation.
\end{abstract}

\smallskip

{\bf  Mathematics Subject Classifications (2010): }  	34A34,  %Nonlinear equations and systems, general
37N05, %Dynamical systems in classical and celestial mechanic
53C15, %General geometric structures on manifolds
 	70F16 %Collisions in celestial mechanics, regularization

\bigskip

{\bf  PACS numbers: } 
02.30.Hq, %Ordinary differential equations
02.40.Yy, 	%Geometric mechanics 
02.40.Hw, 	%Classical differential geometry
02.40.Ky, %Riemannian geometries
45.10.Na %Geometrical and tensorial methods in Classical mechanics of discretesystems
 
\bigskip

\paragraph{Keywords:} 
Sundman transformation, Tangent bundle, regularisation, Jacobi metrics

%\tableofcontents

\section{Introduction} 

An infinitesimal  time reparametrisation, usually called  Sundman transformation  \cite{S13},   was  introduced  when looking for  an analytic solution to the three-body problem. Such transformation 
is very intriguing, at least from a geometric  perspective, but it allowed to find solutions for many different problems in the theory of differential equations and related physical problems. 
For instance it has been very useful in problems of linearisation of differential equations or to  regularise some  equations of motion and avoid collision singularities. More generally,
 to transform a given equation into another of some appropriate form. 
 
 Even if in the early days of
 the beginning of the 
 nineteenth century the
methods developed for studying differential equations  were  of  an
{\sl ad hoc} character, after the pioneer work by  Lie on symmetry methods,  systematic approaches have been 
developed, most of them based on a  geometric theory of differential equations and dynamical systems.
The geometric approach is intrinsic and the results do not depend on a particular choice of coordinates and may be generalised to infinite
dimensional systems (with some topological difficulties). 
 
Within this approach, an autonomous system of  first order differential equations is replaced 
by a vector field $X$ on a differentiable manifold $M$, the system being used to compute, in a local coordinate 
system, the integral curves of the vector field. But the time is not explicitly appearing in the expression of  the vector field 
and only appears as the parameter of such integral curves, and therefore the geometric interpretation of an infinitesimal change of time  is not clear. It will be shown that 
the infinitesimal time reparametrisation can then simply be understood as a change of the 
dynamical vector field $X$, replacing it by a conformally related  one, $f\, X$, where $f$ is a nonvanishing real  function on the  manifold $M$.

The existence of compatible geometric structures on $M$ provided by  special tensor fields has been shown to be very efficient to 
establish and study hidden properties of the given system of differential 
equations and its solutions. This is the main reason for the usefulness of Sundman transformations, because the tensor fields invariant under $f\, X$ are not, in general, invariant under $X$.

The aim of this paper is to investigate  from a geometric perspective the meaning of such  infinitesimal time reparametrisation, to relate it with changes in the vector field
 describing the dynamical system, as well as  to point out many of its applications.
 Section 2 is devoted to first recall  the classical Sundman transformation and, as an example of its applications, to show its use in the linearisation of the Kepler problem, 
 and then to introduce,  from a geometric perspective,  a concept of generalised  Sundman transformation for systems of first-order differential equations.  
Section 3 points out some possible applications in mathematics and classical mechanics. 

The corresponding generalisation of  Sundman transformation  for   systems of second-order differential equations is not so easy and we restrict ourselves in this article
 to the case of those derivable from a variational principle. Therefore,  as we also aim to study this  generalised  Sundman transformation in the  framework of  Riemannian manifolds, 
in order to the paper be self contained,   we recall  in Section 4 the main definitions and properties to be used in the geometric study of  such transformation, and in particular 
the conformal equivalence relation in the set of Riemanniant metrics is explicitly given. A brief overview of symplectic geometry and Lagrangian formalism  is presented
 to introduce some additional notation.  

The rest of the paper is devoted to different applications in geometry and mechanical systems. In Section 5, we describe geodesics  and free motions in a Riemannian 
manifold and the relation with geodesic vector fields and the effect of 
a  Sundman transformation on the set of geodesic curves,  while Section 6 is devoted to mechanical and Newtonian systems in a Riemannian manifold. In both cases we solve the question of the relation between two vector fields whose integral curves are related by a time reparametrisation.

In Section 7 we take a different approach. Instead of changing the vector 
field, we change the metric on the manifold to a conformally related one in order to obtain some specific properties of the vector field of a mechanical system or
 its trajectories. In particular we obtain  in a new fashion, related to a Sundman transformation, the well known 
Jacobi metric. Different relations between the Jacobi metric and the Hamilton-Jacobi equation 
for a Newtonian system are briefly commented in Section 8.

As usually in this kind of geometric papers, all the manifolds and mappings are assumed to be of ${\mathcal C}^{\infty}$ class.

\section{Sundman transformation} 

The classical Sundman transformation \cite{S13} introduced to regularise 
the equations of motion and avoid collision singularities (see also \cite{LC20}), which had previously been used by Levi-Civita \cite{LC04,LC06},  
was shown to be useful in many other situations and can be generalised and extended to other more general cases. For instance the classical paper by Bohlin \cite{B11} used a similar relation to define the Keplerian anomaly. Moreover, 
 Sundman transformation can be used in the study of  linearisation of differential equations  \cite{du94} and   in numerical solution of systems of differential equations (see e.g. \cite{BI,CHL,N76,Euler97,Eulers04}).
   
 The classical Sundman transformation is an infinitesimal scaling of time from the time $t$ to a new fictitious time $\tau$ given by
 \begin{equation}   
    dt= r\, d\tau,  \label{Sundman}
     \end{equation} where $r$ is the radial distance in the plane,  which was later on generalised to $dt= c\,r^\alpha\, d\tau$, where $c\in\mathbb{R}$ and $\alpha$ is a positive constant \cite{N76}, or more generally to $dt= f(r)\, d\tau$ \cite{B85,FS,SB}. 

\subsection{An illustrative example}
      
We next present an explicit example of application of the transformation (\ref{Sundman})  in the Kepler problem.  It is well known that the motion of a particle under a central force takes place in a plane and we can restrict our study to such a plane. Using the standard polar coordinates the 
Lagrange function for a $m=1$ particle  is given by 
$$L(r,\theta)=\frac 12 (\dot r^2+r^2\,\dot\theta^2)-V(r), \quad r>0,
$$
and as the angular variable $\theta$ is cyclic,  the Euler--Lagrange equation for  such variable shows that the corresponding (angular) momentum $\ell=r^2\,\dot\theta$   is constant.
The other Euler--Lagrange equation, $\ddot r=r\,\dot\theta^2-V'(r)$, reduces,  using the constant of motion $\ell$, to the equation of motion of the particle in $(0,\infty)$  under the action of the reduced potential 
$\mathcal{V}(r)=V(r)+\ell^2/(2r^2)$, i.e. in the case of Coulomb--Kepler problem for which $V(r)=-k/r$, the radial equation of motion is
\begin{equation}
\ddot r=\frac {\ell^2}{r^3}-V'(r)=-\mathcal{V}'(r)=\frac {\ell^2}{r^3}-\frac k{r^2}. \label{radialKepler}
\end{equation}

    The conserved energy is given by 
    \begin{equation}
	E=\frac 12 \dot r^2+\frac{\ell^2}{2r^2}-\frac k{r},\label{KEnergy}
	\end{equation}
	and when we introduce the eccentric anomaly parameter $\tau$ by the classical Sundman transformation (\ref{Sundman}), then we get 
	 \begin{equation}
	\frac d{dt}=\frac 1r  \frac d{d\tau},\qquad \frac {d^2}{dt^2}=\frac 1{r^2}\frac {d^2}{d\tau^2}-\frac{r'}{r^3}  \frac d{d\tau}, %\label{dtdtau}
	\end{equation}
	where derivatives with respect to $\tau$ of a function $f$  are denoted by $f^\prime$ instead of $\dot f$. The differential equation (\ref{radialKepler}) becomes 
	 \begin{equation}   
	\frac 1{r^2} r''-\frac 1{r^3}r^{\prime 2}=-\mathcal{V}'(r)=\frac {\ell^2}{r^3}-\frac k{r^2},
  \label{radialKeplertau}
     \end{equation}
     and simplifying 
     \begin{equation} 
      r''=\frac 1r r^{\prime 2}+\frac {\ell^2}{r}-k.  \label{radialKeplertau2}
      \end{equation}
	
But  the expression of the energy (\ref{KEnergy}) can now  be rewritten as 
	 \begin{equation}
	2r^2E={r'}^2+{\ell^2}-2k{r},\label{KEnergy2}
	\end{equation}
	and therefore we see that for motions with a fixed energy $E$, the equation (\ref{radialKeplertau2}) reduces to 
	  \begin{equation}
	r''=2r E +k,\label{linearKeq}
	\end{equation}
which is an inhomogeneous linear second order differential  equation with constant coefficients, whose 
general solution is easily found. 
This shows that Sundman transformation (\ref{Sundman}) provides a linearisation of the motion  equation for the Kepler problem with a fixed energy $E$.

For instance, when the energy is negative and the angular momentum is different from zero, the solutions are ellipses  given by \cite{KOP}
\begin{equation}
	r(\tau)=A(1-e\, \cos(\omega\tau)),\quad \omega=\sqrt{2|E|}, \label{ellipsessol}
	\end{equation}
	where $A$ is the major semiaxis and 
	\begin{equation}
	 t=A\left(\tau-\frac e \omega\,\sin(\omega\tau)\right). \label{ttauK}
    \end{equation}

     \subsection{Generalised Sundman transformation and geometric approach}\label{gensund}
     The important point is that even if this classical  Sundman  transformation (\ref{Sundman}) was carried out for the second order differential equations of motion, it 
admits a generalisation to the case of systems
    of first-order differential equations and, moreover, its geometric interpretation is more  clear. Recall that a
      second-order differential equation vector field in $Q$ can be seen (see Subsection \ref{lagsystems}) as a 
    particular case of vector fields on $TQ$, and therefore it is enough to consider, for the time being, Sundman  transformations for autonomous systems of first-order differential equations.
    
    As indicated in \cite{PRV},  given  an autonomous system of first order differential equations 
\begin{equation}
   \frac{dx^i}{dt}=X^i(x^1,\ldots,x^n), \quad i=1,\ldots,n,\label{autsyst}
   \end{equation}
   we can consider  the generalisation of Sundman transformation defined by 
   \begin{equation}
    dt=f(x^1,\ldots,x^n)\ d\tau,  \quad f(x^1,\ldots,x^n)>0, \label{genSundman}
   \end{equation}
   and then (\ref{autsyst}) becomes 
   \begin{equation}
   \frac{dx^i}{d\tau}=f(x^1,\ldots,x^n)\ X^i(x),\quad i=1,\ldots,n.\label{newautsys}
   \end{equation}
   
   Let us first remark that when each one of  the integral curves of a vector field $X$ is arbitrarily reparametrised we obtain a new family of curves which may be, or not, the integral curves of a vector field $Y$. In the affirmative case, as the two vector fields have the same local constants of motion, they generate the same 1-dimensional distribution and, at least locally, there
    exists a nonvanishing function $h$ such that $Y=h\, X$. Let us prove that this is the case if we consider the reparametrisation defined by a Sundman transformation and then  $h$ coincides with the  
    the function $f$ defining the transformation (\ref{genSundman}).
   
     In fact,  if $\gamma(t)$ is a given curve and we carry out the reparametrisation for which  the new parameter  $\tau$ is defined by the relation (\ref{genSundman}) written as 
\begin{equation} 
\frac{d\tau}{dt}=\frac 1{ f(\gamma(t))},\label{genSundman2}
\end{equation} 
we obtain the reparametrised curve $\bar\gamma(\tau)$ such that  $\bar\gamma(\tau(t))=\gamma(t)$ and then
$$ \frac{d\gamma}{dt}=\frac{d\bar\gamma}{d\tau}\,\frac{d\tau}{dt}=\frac 1{ f(\gamma(t))}\frac{d\bar\gamma}{d\tau},
$$
and consequently, if the curve  $\gamma(t)$   is an integral curve of $X$, i.e. $d\gamma/dt=X_{\gamma}$,  then 
$$
\frac{d\bar\gamma}{d\tau}= f(\gamma(t)) \frac{d\gamma}{dt}=f(\gamma(t))\,X_{ \gamma(t)}=(f\,X)_{\gamma(t)}=(f\,X)_{\bar \gamma(\tau)}\,,
$$
i.e. the curve $\bar\gamma$ is an integral curve of the vector field $f\, 
X$.

   From the geometric viewpoint the solutions of the system  (\ref{autsyst}) provide us the integral curves of the vector field $X=X^i(x^1,\ldots,x^n)\partial/\partial {x^i}$ and 
   then the solutions of the new system  (\ref{newautsys}) provide the integral curves of the vector field $f\, X$. In other words, the effect of 
   the generalised  Sundman transformation (\ref{genSundman}) is the replacement by the vector field $f\, X$ instead of $X$ \cite{CCJM}. But  the integral curves of the vector fields 
$X\in \mathfrak{X}(M)$  and $f\,\,X\in \mathfrak{X}(M)$ coincide up to respective reparametrisations, because both vector fields  have the same local constants of motion.

Equivalently,   the reparametrisation defined by Sundman transformation (\ref{genSundman2}) should be carried out for each orbit, i.e. as pointed out in  \cite{MM11}, if $x(t)$ is a solution of (\ref{autsyst}), then we consider the reparametrisation defined by 
$$\tau(t)=\int_0^t \frac 1{f(x(\zeta))}\ d\zeta , $$
and the inverse expression $t=\varphi(\tau)$, and then $x(\varphi(\tau))$ is a solution of (\ref{newautsys}). 

It is noteworthy that the `velocity' with respect to the new time is different and so the new velocity $\bar v$ is  related to the old one by $\bar v^i=f\, v^i$, as a
 consequence of (\ref{genSundman2}). 

The reinterpretation of this `infinitesimal time scaling' was used in \cite{CIL}  to deal with the theory described by Bond and Janin in \cite{BJ} 
 in satellite theory.
 By appropriately selecting the function $f$, a problem with singular solutions
 can be transformed into a related one with globally defined solutions  in terms of 
the new time variable.

 \section{Applications in mathematics and classical mechanics}
 
 Let us first remark that even if  the vector fields $X$ and $f\,X$ on a manifold $M$, with $f$ a nonvanishing function, usually called conformally related vector fields,  have the same local constants of motion, the same  property does not hold for general tensor fields $T$, because $\mathcal{L}_{f\,X}T\ne f\, \mathcal{L}_{X}T$  and this fact is quite important because 
 it provides us a way for obtaining,  from a vector field $X$ that does 
not preserve $T$,  vector fields under which a given  
tensor field $T$ is invariant. 

We have shown before how the classical  Sundman transformation (\ref{Sundman})  provides a
  method to linearise the equations of motion for a given energy in the Kepler--Coulomb problem. Recall that a manifold $M$ can be endowed with a linear structure (see \cite{CIMM}, Chapter 3)  if there exists a complete 
vector field $\Delta$, playing the r\^ole  of Liouville vector field, 
  with only one non-degenerate critical point  and   such that $\mathcal{F}_\Delta^{(0)}=\mathbb{R}$ and  $\mathcal{F}_\Delta^{(1)}$ separates derivations, where $\mathcal{F}_\Delta^{(k)}$ 
denotes the set of functions on the manifold $M$ defined by 
$$\mathcal{F}_\Delta^{(k)}=\{f\in \mathcal{F}(M)\mid \Delta f=k\,f\}, 
\quad  k\in \mathbb{N}.$$ 
Linear functions (with respect to such a linear structure) are those $f\in \mathcal{F}(M)$ satisfying that $ \mathcal{L}_{\Delta}f=f$, that is, the elements of $\mathcal{F}_\Delta^{(1)}$, and linear vector fields $X\in \mathfrak{X}(M)$  are those preserving the $\mathbb{R}$-linear subspace of linear functions, that is, satisfying 
$ \mathcal{L}_{X}(\mathcal{F}_\Delta^{(1)})\subset\mathcal{F}_\Delta^{(1)}$, or what is equivalent, 
$ \mathcal{L}_{X}\Delta=[X,\Delta ]=0$.

Then, given a nonlinear vector field 
$X$ such that  $ \mathcal{L}_{X}\Delta\ne 0$,  it may exists a positive function $f\in C^\infty(M)$ such that $\mathcal{L}_{f\,X}\Delta=0$, because 
 $\mathcal{L}_{f\,X}\Delta=[f\, X,\Delta]=f\,[X,\Delta ]- \Delta(f)\, 
X$. This happens when there exists a function $h$ such  that  $[X,\Delta ]=h\,X $ and 
 we choose the function   $f$ in such a way that $h=\Delta(\log f)$. 

In the  case of $M$ being a $n$-dimensional manifold,  a chart $(U,\varphi)$ of $M$ identifies $U$ 
with an open set of $\mathbb{R}^n$, and this one,  as a linear space,  is 
endowed with the complete vector field $\Delta={\displaystyle \sum_{i=1}^n}x^i\,\partial/\partial x^i$, where $x^i$ are the usual coordinates, and therefore, if $X={\displaystyle \sum_{i=1}^n}X^i(x)\, \partial/\partial x^i$ is the coordinate expression of the vector field $X$ in the
 mentioned chart, then we can say that the vector  field $X$ is linear  in this chart when  $[\Delta, X] =0$, and taking into account that 
\begin{equation}[\Delta, X] = \left[\sum_{i=1}^n x^i\pd{}{x^i}, \sum_{j=1}^n X^j\pd{}{x^j} \right]=  \sum_{k=1}^n\left(\sum_{i=1}^nx^i\pd{X^k}{x^i}-X^k\right)\pd{}{x^k},\label{DeltaX}
\end{equation}
 we see from $[\Delta, X] =0$ that the components of the vector field $X$ in the given chart are homogeneous of order one functions, i.e.  linear functions, $X^i(x)={\displaystyle\sum_{j=1}^n}A^i\,_j\, x^j$,  where $A^i\,_j\in\mathbb{R}$, and then their integral curves are given by the  solutions of the linear system $\dot x^i={\displaystyle\sum_{j=1}^n}A^i\,_j\, x^j$, $i=1,\ldots,n$. The vector fields $X$  whose  integral curves are given by solutions of inhomogeneous linear systems $\dot x^i={\displaystyle\sum_{j=1}^n}A^i\,_j\, x^j+B^i$, $i=1,\ldots,n$, where $B^i\in\mathbb{R}$, are  the sum of  a linear vector field $X_0$  and another one $X_{-1}$ with components that are homogeneous of degree zero, i.e. constants. Of course as $[\Delta, X_0]=0$ and $[\Delta, X_{-1}]= -X_{-1}$ we have that
$$[\Delta, X]= [\Delta, X_0+X_{-1}]= [\Delta, X_0]+[\Delta, X_{-1}]= -X_{-1},
$$
and hence,
$$[\Delta ,[\Delta ,X]]=-[\Delta,  X_{-1}]= X_{-1}=   -[\Delta, X].
$$

Conversely, if this last relation holds, then having in mind (\ref{DeltaX}) we see that 
$$[\Delta,[\Delta,X]]=\sum_{i=1}^n\left(\Delta(\Delta(X^i)-X^i)-\Delta(X^i)-X^i\right )\pd{}{x^i},$$
and therefore, $[\Delta ,[\Delta ,X]]=  -[\Delta, X]$ if and only if, for each index $i=1,\ldots,n$,  $\Delta^2(X^i)-\Delta(X^i)=0$, i.e. $\Delta(X^i)-X^i$ is a  homogeneous  function of order zero,
because there exist  constants $B^i$ such 
$$
\sum_{j=1}^nx^j\pd{X^i}{x^j}-X^i=- B^i, 
$$
from where we find that $X^i={\displaystyle \sum_{j=1}^n}A^i\,_jx^j+B^i$,  for each index $i=1,\ldots,n$. Without using local coordinates, we observe that $[\Delta, [\Delta, X]]=-[\Delta, X]$ implies that the components of the 
vector field  $[\Delta, X]$ are homogeneous functions of order zero, i.e. is a vector field with constant components, i.e., $X$ is if the form $X=X_0+X_{-1}$.

We can summarise the preceding results as follows:

\begin{theorem}
Let $M$ be a differentiable manifold endowed with a linear structure $\Delta$. If  $X$ is a vector field in $M$, then:
\begin{enumerate}
\item $\mathcal{L}_{\Delta}X=h\, X$, for some function $h$, if and only if there exists a nonvanishing positive function $f$ such that the vector field $fX$ satisfies $\mathcal{L}_{\Delta}(fX)=0$. 

\item $\mathcal{L}_{\Delta}X=0$ if and only if for any coordinate system $(U,\varphi)$, where  $\varphi=(x^{1},\ldots,x^{n})$,  such that
 $\left.\Delta\right|_{U}={\displaystyle \sum_{i=1}^nx^{i}\frac{\partial}{\partial x^{i}}}$, the local expression of $X$ is linear in $x^{i}$, that is, 
$$\left. X\right|_{U}=\sum_{i,j=1}^nA^i\,_j\, x^{j}\frac{\partial}{\partial x^{i}}, \quad A^i\,_j\in\mathbb{R}.$$

\item $\mathcal{L}_{\Delta}^{2}X=-\mathcal{L}_{\Delta}X$ if and only if for any coordinate system $(U,\varphi=(x^{1},\ldots,x^{n}))$ such that $\left.\Delta\right|_{U}={\displaystyle \sum_{i=1}^n }x^{i}\frac{\partial}{\partial x^{i}}$, the local expressi\'on of $X$ is affine in $x^{i}$, that is, 
$$\left. X\right|_{U}=\sum_{i=1}^n\left(\sum_{j=1}^nA^i\,_j\, x^{j}+B^i\right) \pd{}{x^i}, \quad A^i\,_j,B^i\in\mathbb{R} .$$
\end{enumerate}
\end{theorem}
\hfill$\Box$

\noindent {\bf Comment}: This theorem gives us a characterisation of the vector fields $X$ that are linearisable with respect to a general linear structure $\Delta$ in a manifold $M$, by means of a general Sundman transformation.

\bigskip 
  
   As another instance of application of Sundman transformations, if   $(M,\Omega)$ is an oriented manifold, i.e. $\Omega$ is a volume form in $M$,   and a vector field $X\in \mathfrak{X}(M)$   is such that  $\mathcal{L}_X\Omega\ne0$, there may be a positive function   $f\in C^\infty(M)$ such that $\mathcal{L}_{f\,X}\Omega=0$: these functions are called   Jacobi multipliers \cite{JCB09} and play a relevant r\^ole  in the process  of integrability by quadratures.    Note that as $\mathcal{L}_{f\,X}\Omega=\mathcal{L}_{X}(f\,\Omega)$, the search of such a Jacobi multiplier is 
   equivalent to the determination of a $X$-invariant volume form $f\,\Omega$. For examples of applications of  Jacobi multipliers  in integrability and the inverse problem of mechanics
   see e.g. the recent review paper \cite{CFN21} and references therein.

  Another possibility would be to consider a Riemannian structure on the manifold $M$ given by a non-degenerate symmetric
   two times covariant tensor field $g$ and then, as $\mathcal{L}_{f\,X}g\ne f\,\mathcal{L}_Xg$, some non-Killing vector fields $X$, i.e such that $\mathcal{L}_Xg\ne 0$,  can give rise to  Killing ones by just multiplication by a convenient function $f$. 
   
   Similarly, given a closed 2-form on a manifold, for instance a symplectic form $\omega$, 
 that may be  not invariant under the vector field $X$,  i.e.  $\mathcal{L}_{X}\omega\ne 0$, if there is 
 a positive  function $f$ such that  $\mathcal{L}_{f\,X}\omega=0$, i.e. 
$f\,i(X)\omega$ is a closed form,   then, when  the non-degeneracy condition $df\wedge \omega=0$ is satisfied,  the 
2-form $f\,\omega$ is symplectic and  the vector field $X$ is locally-Hamiltonian with respect to such symplectic form $f\,\omega$.  Recall that in the case of $X$ being also locally-Hamiltonian with respect to the original symplectic form  $\omega$, such function $f$ must be a constant of motion for $X$, $Xf=0$ (see \cite{CI}).
 
 The same can be said about skew-symmetric
   two times contravariant tensor fields, i.e. bivector fields $\Lambda$, 
 and a particularly interesting case is that of Poisson structures, i.e. such that $[\Lambda,\Lambda]_{\textrm{S}}=0$, where  $[\cdot,\cdot]_{\textrm{S}}$ denotes the Schouten bracket \cite{koszul, L77}. Recall that in this case 
   $$\{f,g\}=\Lambda(df,dg),\quad f,g\in C^\infty(M),
   $$
   defines a Poisson bracket and vector fields $X_f$ on $M$ of the form $X_fg=\{g,f\}$ are said to be Hamiltonian vector fields. Then given a vector field $X$ and a Poisson structure $\Lambda$ sometimes there exists a 
function $f$ such that $fX$ is Hamiltonian with respect to the Poisson structure $\Lambda$. In this way an infinitesimal Sundman time-reparametrisation
   can transform a given vector field into a Hamiltonian one. This  is usually called Hamiltonization process \cite{BBM15a, BBM16}.

 \section{Digression on some geometric complements}   
 
  The  applications of  the Sundman transformation for systems of second-order differential equations we will develop in next sections cover different topics, going from minimal length curves to systems of a mechanical type  and from the geodesic vector fields to the Jacobi metric. This is why we need to extend our standpoint and formulate the Sundman transformation in a different framework, that of a Riemannian manifold.
  
  We begin with a short review of the concepts of Riemannian geometry we will need, some comments on conformally related  metrics and some ideas on the tangent bundle and second-order  differential equation vector fields in the case of Lagrangian systems and   systems  of a mechanical type. The remaining sections are devoted to a detailed study of different applications using these topics.

  \subsection{A quick survey of Riemannian geometry}
  
In order to the paper be self-contained, this Subsection is devoted to recall well known concepts and properties in geometry  and also to establish the notation to be used.  The results can be found in many classical books   on Riemann geometry (see   e.g. \cite{L18} and \cite{C08} for more details).

 A (pseudo-)Riemann manifold $(M,g)$ is a pair given by a differentiable manifold $M$, $\dim M=n$, and a non-degenerate  symmetric
 two times covariant tensor field  $g$ on $M$. Nondegeneracy means that the map $\widehat g:TM\to T^*M$ from the tangent bundle, $\tau_M:TM\to  M$, 
 to the cotangent bundle, $\pi_M:T^*M\to M$,  defined by $\<\widehat g(v),w>=g(v,w)$, where  $v,w\in T_xM$, is a regular vector bundle map, that 
is,  a vector bundle isomorphism. As the map $\,\widehat g\,$ is a fibred 
map
  over the identity on $M$, it  induces the corresponding map between the 
$C^\infty(M)$-modules of sections of the tangent and cotangent bundles, to be denoted by the
   same letter $\widehat g:\mathfrak{X}(M)\to \bigwedge^1(M)$: $\<\widehat g(X),Y>=g(X,Y)$. Given a local chart $(U,q^1,\ldots,q^n)$ on $M$ we can  
consider the coordinate basis of $\mathfrak{X}(U)$ usually denoted $\{\partial/\partial q^j\mid j=1,\ldots ,n\}$ and its dual basis for $\bigwedge^1(U)$, $\{dq^j\mid j=1,\ldots ,n\}$. Then a  vector $v$ in a point $q\in U$  is $v={\displaystyle\sum_{j=1}^n}v^j\,(\partial/\partial q^j)_{|q}$ and a covector $\zeta$ in such a point  is $\zeta={\displaystyle\sum_{j=1}^n}p_j\,(dq^j)_{|q}$, with  $v^j=\<dq^j,v>$ and $p_j=\<\zeta,\partial/\partial q^j>$ being the usual velocities and momenta. The local expression for 
$g$ in the open set $U$ is 
\begin{equation}
g=\sum_{i,j=1}^ng_{ij}(q)\,dq^i\otimes dq^j, \quad g_{ij}(q)=g\left(\pd{}{q^i},\pd{}{q^j}\right),\quad i,j=1,\ldots,\dim M, \label{Rstr}
\end{equation}
and therefore the arc-length of a curve $\gamma$  in $M$, $\gamma(t)=(q^1(t),\ldots, q^n(t))$, between the points $\gamma(t_1)$ and $\gamma(t_2)$    is given by 
\begin{equation}
s(\gamma)=\int_{t_1}^{t_2} \sqrt{g(\dot\gamma,\dot\gamma)}\, dt=\int_{t_1}^{t_2} \sqrt{{\displaystyle \sum_{i,j=1}^n}g_{ij}(q)\dot q^i\dot q^j}\ dt ,\label{length}
\end{equation}
i.e. the classical  local expression for the arc-length    $ds$  is given 
  by 
\begin{equation}
ds^2 =\sum_{i,j=1}^ng_{ij}(q)\,dq^idq^j.\label{dsg}
\end{equation}

Given $q_1,q_2\in M$, {\sl the extremal length curves} in $(M,g)$ from $q_{1}$ to $q_{2}$ are the curves $\gamma:[t_1,t_2]\to M, \gamma(t_1)=q_1,\gamma(t_2)=q_2$, such that the integral (\ref{length}) is extremal among all the curves going from $q_{1}$ to $q_{2}$. Such curves will be studied in  Subsection 5.1.

Recall that a linear connection $\nabla$ in a manifold $M$  is a  map 
 $\nabla:\mathfrak{X}(M)\times
\mathfrak{X}(M)\to \mathfrak{X}(M)$, such that if $X,Y,Z\in\mathfrak{X}(M)$, $f\in C^\infty(M)$ and  $\nabla _XY$ denotes 
   $\nabla _XY=\nabla(X,Y)$:
$$\begin{array}{rl} {\text{i)}}&\nabla_{X+Y}Z=\nabla_XZ+\nabla_YZ,\\
{\text{ii)}}&\nabla_X(Y+Z)=\nabla_XY+\nabla_XZ,\\
{\text{iii)}}&\nabla_{fX}Y=f\, \nabla_XY,\\
{\text{iv)}}&\nabla_X(fY)=(Xf)Y+f\, \nabla_XY.
\end{array}
$$

In a local coordinate system the connection symbols 
$\Gamma^i_{jk}$ are defined by 
\begin{equation}
\nabla_{\pd{}{x^i}} \left(\pd {}{x^j}\right)=\sum_{l=1}^n\Gamma^l_{ji}\,
\pd {}{x^l},\label{connsymbols}
\end{equation}
while $\nabla_XY$ is given by 
\begin{equation}
\nabla_XY=\nabla_{\sum_{i=1}^nX^i\pd{}{x^i}}\left (\sum_{j=1}^nY^j \pd {}{x^j}\right)=\sum_{i,k=1}^n
X^i\left (\pd {Y^k}{x^i}+\sum_{j=1}^nY^j\Gamma^k_{ji} \right)\pd {}{x^k}.\label{deltaXY}
\end{equation}
Moreover, each linear connection has associated a (1,2)  tensor $\mathbb{T}$, called torsion tensor,  which is skewsymmetric in the  two last indices and it is
defined as follows
\begin{equation}
\mathbb{T}(X,Y)=\nabla_XY-\nabla_YX-[X,Y], \label{torsiont}
\end{equation}
and a (1,3) tensor field called curvature tensor defined by 
\begin{equation}
R(X,Y)Z=\nabla_X(\nabla_YZ)-\nabla_Y(\nabla_XZ)-\nabla_{[X,Y]}Z.\label{curvatt}
\end{equation}

A Riemann manifold $(M,g)$ is endowed with 
 a uniquely determined torsionless metric connection $\nabla$, called the 
Levi--Civita connection. By
 metric we understand that if   $X,Y,Z\in \mathfrak{X}(M)$, then
  \vspace{-2mm}
 \begin{equation}
 \mathcal{L}_X(g(Y,Z))=g(\nabla_XY,Z)+g(Y,\nabla_XZ),\label{metriccon}
 \end{equation}
 and by torsionless we mean that the corresponding torsion tensor $\mathbb{T}$ is null, hence
 \vspace{-3mm}
    \begin{equation} 
\nabla_XY-\nabla_YX=[X,Y].\label{torsionless}
 \end{equation}
 Such a Levi-Civita connection is given by the so-called
 Koszul formula:  
    \begin{equation} 
    \begin{array}{rcl} 2g(\nabla_XY,Z)&=&\mathcal{L}_X(g(Y,Z))+\mathcal{L}_Y(g(Z,X))-\mathcal{L}_Z(g(X,Y))\\&&\\&
-&g(X,[Y,Z])+g(Y,[Z,X])+g(Z,[X,Y ]),\end{array} \label{koszul}
 \end{equation} 
for every $X,Y,Z\in \mathfrak{X}(M)$.
In  terms of local coordinates on $M$  Koszul formula gives   $$2g\left(\nabla_{\partial/\partial
x^i}\left(\pd{}{x^j}\right),\pd{}{x^k}\right)=\pd{g_{jk}}{x^i}
+\pd{g_{ik}}{x^j}-\pd{g_{ij}}{x^k}, $$ and using the definition of connection symbols  
(usually called second class Christoffel symbols) 
$$g\left(\nabla_{\partial/\partial
x^i}\left(\pd{}{x^j}\right),\pd{}{x^k}\right)=\sum_{l=1}^n\Gamma_{ji} ^l\,g_{lk}. $$ Hence,
the second class Christoffel symbols are given by
\begin{equation}
\Gamma^i_{jk}=\frac 12 \sum_{l=1}^n g^{il} \left(\pd{g_{lj}}{q^k}+\pd{g_{lk}}{q^j}-\pd{g_{jk}}{q^l}\right),\label{Gijk}
\end{equation}
where ${\displaystyle  \sum_{j=1}^n}g^{ij} g_{jk}=\delta^i_k$.

On the other hand, the vanishing of the torsion tensor is equivalent to the following symmetry property of  Christoffel symbols:  $\Gamma^i_{jk}=\Gamma^i_{kj}$.

As indicated above,  these symbols $\Gamma^i_{jk}$ locally determine $\nabla_XY$ for every $X,Y,$ by making use of  (\ref{deltaXY}). The remarkable fact (see e.g. \cite{L18,C08} for details)  is that as given a vector field $Y\in\mathfrak{X}(M)$  the value of $\nabla_XY$ at a point $x\in M$,  $(\nabla_XY)(x)$,  only depends, with respect to $X$, on the value of $X$ in such a point, then for each $v\in T_xM$ we can define $\nabla_vY=(\nabla_XY)(x)$, where $X$ is any vector field $X\in\mathfrak{X}(M)$ such that $X(x)=v$, and then this allows us to introduce the concept of parallelism of a vector field along a curve  as follows:   A vector field along a curve $\gamma:I
 \to M$, $Y\in  \mathfrak{X}(\gamma)$, is parallel along $\gamma$ if $\nabla_{\dot \gamma(t)} \bar Y=0$, for all $t\in I$, where $\bar Y\in\mathfrak{X}(M)$ is an arbitrary  extension to $M$ of $Y$.

An example of a vector field along  a curve $\gamma$ in $M$ is given by its velocity vector field $\dot \gamma\in \mathfrak{X}(\gamma)$, and the curves whose velocity vector field  is parallel along the curve, $\nabla_{\dot \gamma(t)}\dot\gamma(t)=0$,  are called geodesics. In particular in a Riemann  manifold $(M,g)$ the connection to be considered is its Levi-Civita connection
 and using this connection $\nabla$, the geodesics in $(M,g)$ are the curves $\gamma$  in $M$ 
satisfying the equation  $\nabla_{\dot \gamma}\,\dot \gamma=0$, where  $\dot \gamma$ is the vector field 
along $\gamma$ given by the velocity of the curve at each point. They are uniquely defined from the Riemann metric $g$  through  the Levi-Civita 
connection $\nabla$, which is uniquely defined  by  $g$. Their geometric properties can be studied through the so called second order geodesic vector field defined by $g$ and we will see below the relation with extremal 
length curves through an appropriate Sundman transformation. 

For a curve parametrised by its arc-length $s$, $\gamma(s)=(q^{1}(s),\ldots,q^{n}(s))$, as we have
$$
\nabla_{\dot \gamma}\,\dot \gamma=\sum_{l=1}^n\left(\sum_{j=1}^n\dot q^{j}\frac{\partial \dot q^{l}}{\partial q^{j}}+\sum_{j,k=1}^n
\Gamma^l_{jk}\dot q^{j}\dot q^{k}\right) \frac{\partial }{\partial q^{l}}\circ{\gamma}=\sum_{l=1}^n\left(\ddot q^{l}+\sum_{j,k=1}^n\Gamma^l_{jk}\dot q^{j}\dot q^{k}\right) \frac{\partial }{\partial q^{l}}\circ\gamma\, ,
$$
the local equation of the geodesic lines, $\nabla_{\dot \gamma}\,\dot \gamma=0$, is the second order differential equation
\begin{equation}\label{geodeq}
\ddot q^{i}+\sum_{j,k=1}^n\Gamma^i_{jk}\dot q^{j}\dot q^{k}=0\, ,\quad l=1,\ldots,n\,.
\end{equation}
See Subsection 5.1 for another approach to this equation.

Consequently, the geodesic curves are the projection on the base manifold $M$ of the integral curves of the second order geodesic vector field $\Gamma\in\mathfrak{X}(TM)$
whose local expression is
\begin{equation}\label{2ogeodvf}
\Gamma= \sum_{i=1}^n\left(v^i\pd{}{q^i}-\left(\sum_{j,k=1}^n\Gamma^i_{jk}v^{j}v^{k}\right)\pd{}{v^i}\right).
\end{equation}
 
\subsection{Conformal metrics}
We have seen in Section \ref{gensund} that the vector fields $X$ and $fX$ on a manifold $M$
have integral curves related by a Sundman transformation. Other interesting problems related to Sundman transformation for second-order differential equation vector fields  use 
conformal Riemannian metrics to study dynamical systems. Two metrics, $g$ and $\bar g$, are conformally related if there exists a function $\varphi$ such that
 $\bar g=\exp(2\varphi)\, g$. This establishes an equivalence relation in the set of metrics.   A conformal structure is an equivalence class of metrics. The   covariant derivatives with respect to both metrics  $g$ and $\bar g$
 are related by:
\begin{equation}
\begin{array}{rcl}
\bar\nabla_XY&=&  
\nabla_XY+(\mathcal{L}_X\varphi)Y+(\mathcal{L}_Y\varphi)X-g(X,Y)\,{\textrm{grad}}_g\varphi\\&&\\& =&\nabla_XY+(\mathcal{L}_X\varphi)Y+(\mathcal{L}_Y\varphi)X-\bar g(X,Y)\,{\textrm{grad}}_{\bar g}\varphi\, .
\end{array}\label{barnabla}
\end{equation}
To obtain these formulas, consider the Koszul formula for the connection $\bar\nabla$  that gives 
\begin{equation}
\begin{array}{rcl}
e^{2\varphi}g(\bar\nabla_XY,Z)&=&e^{2\varphi}\left[\mathcal{L}_Xg(Y,Z)+\mathcal{L}_Yg(Z,X)-\mathcal{L}_Zg(X,Y)+2g(Y,Z)X(\varphi)+2g(Z,X)Y(\varphi)\right.\\&&\\&-&\left.2g(X,Y)Z(\varphi) -g(X,[Y,Z])+g(Y,[Z,X])+g(Z,[X,Y ])\right],
\end{array}
\end{equation}
and simplifying the factor $e^{2\varphi}$, it may be rewritten as
\begin{equation}
\begin{array}{rcl}
2g(\bar\nabla_XY,Z)&=& \mathcal{L}_Xg(Y,Z)+\mathcal{L}_Yg(Z,X)-\mathcal{L}_Zg(X,Y)+2g(Y,Z)X(\varphi)+2g(Z,X)Y(\varphi)\\&&\\&-&2g(X,Y)Z(\varphi)
-g(X,[Y,Z])+g(Y,[Z,X])+g(Z,[X,Y ]) ,
\end{array}
\end{equation}
from where, once again  by simplification, we obtain
\begin{equation}
g(\bar\nabla_XY,Z)=g(\nabla_XY,Z)+g(Y,Z)X(\varphi)+g(Z,X)Y(\varphi)-g(X,Y)Z(\varphi).
\end{equation}
Finally, by making use of 
$$g(X,Y)\mathcal{L}_Z\varphi=g(X,Y) d\varphi(Z)=g(X,Y)g(\textrm{grad}_g \varphi,Z)=g(g(X,Y)\textrm{grad}_g \varphi,Z),
$$
we find 
\begin{equation}
\bar\nabla_XY=  
\nabla_XY+(\mathcal{L}_X\varphi)Y+(\mathcal{L}_Y\varphi)X-g(X,Y)\,{\textrm{grad}}_g\varphi,\label{barnabla1}
\end{equation}
where ${\textrm{grad}}_g$ is the gradient with respect to the metric $g$. The above relation can also be rewritten as  
\begin{equation}
\bar\nabla_XY=  
\nabla_XY+(\mathcal{L}_X\varphi)Y+(\mathcal{L}_Y\varphi)X-\bar g(X,Y)\,{\textrm{grad}}_{\bar g}\varphi.
\label{barnabla2}
\end{equation}
Remark that the relation  $\bar g=\exp(2\varphi)\, g$ 
implies that ${\textrm{grad}}_{\bar g}\varphi =
\exp(-2\varphi)\,{\textrm{grad}}_{g}\varphi $. 

In the case of  $X=Y$   the  expression (\ref{barnabla1}) reduces to
\begin{equation}
\bar\nabla_XX=  
\nabla_XX+2(\mathcal{L}_X\varphi)X-g(X,X){\textrm{grad}}_g\varphi\, .\label{barnabla3}
\end{equation}

 It is also interesting to study the relationship between the corresponding Christoffel symbols to both metrics. The new   Christoffel symbols of the second kind are  \cite{mt14}:
 \begin{equation}
\bar \Gamma^i_{jk}(q)= \Gamma^i_{jk}(q)+\delta^i_j\pd{\varphi}{q^k}+\delta^i_k\pd{\varphi}{q^j}-\sum_{l=1}^ng_{jk }\,g^{il}\pd{\varphi}{q^l}, \quad i,j,k=1,\ldots,\dim M, \label{bGijk} 
\end{equation}
 where $\delta^i_k$ denotes Kronecker delta symbol, because 
$$\bar \Gamma^i_{jk}=\frac 12 e^{-2\varphi} \sum_{l=1}^ng^{il}\left(\pd{}{q^k}(e^{2\varphi}g_{lj})+
\pd{}{q^j}(e^{2\varphi}g_{lk})-\pd{}{q^l}(e^{2\varphi}g_{jk})\right),  \quad i,j,k=1,\ldots,\dim M,
$$     
and using Leibniz rule for derivatives and simplifying terms we arrive to 
(\ref{bGijk}).

There is also an interesting equivalence relation in the set of symmetric linear connections (see e.g. \cite{mt14} and references therein): two such  connections are said to be 
projectively equivalent when their geodesics differ only by a parametrisation. Each equivalence class is characterised by its Thomas symbol. A projective structure is compatible with a conformal structure if there exists a metric $g$ in  its conformal class such  the associated Levi-Civita connection is in the equivalence class defining the  projective structure.
Necessary and sufficient conditions for local compatibility are given in \cite{mt14}.

\subsection{Lagrangian systems}\label{lagsystems}

Given a regular Lagrangian (see later on), $L:TM\to\mathbb{R}$, $L(q,v)$, defined on the velocity phase space, that is,  a function depending on the positions $q$ and velocities $v$, we can obtain the associated dynamical equations in two different ways. In the traditional one we consider the Hamilton principle, that is, we try to determine the curves $q(t)$ 
with fixed end-points making extremal the action defined by: 
$$S(q(t))=\int_{t_1}^{t_2} {L}(q(t),\dot q(t))\, dt.
$$
Then, applying the classical variational calculus, the curves we are looking for  are solutions to the Euler-Lagrange equations associated to the Lagrangian $L$: 
\begin{equation}
\frac d{dt}\left(\pd {L}{\dot q^i}\right)-\pd {L}{q^i}=0,\quad i=1,\ldots,n\,.\label{ELL}
\end{equation}

 But the theory of the above systems can be presented more geometrically by means of the symplectic approach  to Lagrangian formalism (see e.g \cite{CIMM, Cr81, Cr83a}). In this approach, from the Lagrangian function $L$ we construct the Cartan 1-form $\theta_{L}$ given by                    
 $\theta_{L}=d{L}\circ S$, where $S$ is the vertical endomorphism \cite{Cr81, Cr83a} in the tangent bundle $TM$ of the configuration manifold 
$M$. Associated to the 1-form $\theta_{L}$, we have the 2-form $\omega_{L}=-d\theta_{L}$. These forms are  locally given by 
$$
\theta_{L}= \sum_{j=1}^n\frac{\partial L}{\partial v^{j}}dq^j\, ,\qquad
 d\theta_{L}=  \sum_{j=1}^nd\left(\frac{\partial L}{\partial v^{j}}\right)\wedge dq^j= \sum_{i,j=1}^n\left(
 \frac{\partial^{2}L}{\partial q^i\partial v^{j}}dq^i\wedge dq^j+
 \frac{\partial^{2}L}{\partial v^i\partial v^{j}}dv^i\wedge dq^j\right)
 $$
 
If $\omega_L$ is symplectic, we say that the Lagrangian $L$ is \textsl{regular}. Otherwise $L$ is called \textsl{singular}. In the sequel we suppose that $L$ is regular.

The 2-form $\omega_L$ has  the usual  form
\begin{equation}
\widehat \omega_{L}=\begin{pmatrix} A&-W\\W&0\end{pmatrix},\qquad \widehat \omega_{L}^{-1}=\begin{pmatrix} 0&W^{-1}\\-W^{-1}&W^{-1}AW^{-1}\end{pmatrix}\, ,\label{omegaL}
\end{equation}
with the matrices $A$ and $W$ being given by
$$
A_{ij}= \frac{\partial^{2}L}{\partial q^i\partial v^{j}}-\frac{\partial^{2}L}{\partial v^i\partial q^{j}} , \qquad 
W_{ij}= \frac{\partial^{2}L}{\partial v^i\partial v^{j}}, \quad i,j=1,\ldots,n.
$$
The regularity of the Lagrangian depends on the regularity of $\widehat \omega_{L}$  and, consequently, on  the regularity of the matrix $W$.

 On the other side, recall that the energy of the Lagrangian system determined by $L$  is defined by $E_L=\Delta L - L$, where $\Delta\in\mathfrak{X}(TM)$ is the Liouville vector field, generator of dilations along the fibres of $TM$, given by 
  \begin{equation}\Delta f(q,v)=\frac{d}{dt} f(q,e^tv)|_{t=0}, \qquad 
\Delta= \sum_{i=1}^nv^i\frac{\partial}{\partial v^i},\label{Liouvillevf}
  \end{equation}
 for all $(q,v)\in TM$ and $f\in C^\infty(TM)$. 
Hence, as $\Delta({L})= {\displaystyle\sum_{j=1}^n}v^{j}{\partial L}/{\partial v^{j}}$,
  the total energy is $E_{L}= {\displaystyle\sum_{j=1}^n}v^{j}{\partial L}/{\partial v^{j}}-L$. 
  The corresponding associated dynamics when $L$ is regular  is given by the unique dynamical vector field $\Gamma_{L}\in\mathfrak{X}(TM)$ defined 
   by 
\begin{equation}
i(\Gamma_{L})\omega_{L}=dE_{L}, \label{dyneqL}
\end{equation}
and since we know the expression of $E_{L}$,
we see that $\Gamma_{L}$ is of the form
\begin{equation}
\Gamma_{L}=\sum_{i=1}^n\left(v^i\pd{}{q^i}+F^i(q,v)\, \pd{}{v^i}\right), \label{GammaL}
\end{equation}   
where
\begin{equation}
F^i(q,v)=-\sum_{j,l=1}^nW^{ij}A_{lj}v^l+\sum_{j=1}^nW^{ij}\frac{\partial L}{\partial q^{j}}\, , \label{GammaLcoor}
\end{equation}
with ${\displaystyle\sum_{j=1}^n}W^{ij} W_{jk}=\delta^i_k$. 

The uniqueness  of  $\Gamma_{L}$ satisfying equation (\ref{dyneqL}) is a consequence of the assumed  regularity of the 2-form $\omega_L$.

The dynamical vector field $\Gamma_{L}$ is a second order differential equation and the corresponding system of differential equations is
\begin{equation}
\left\{
\begin{array}{rcl}
\dfrac{d q^{j}}{dt}&=&v^{j}\\
\dfrac{d v^{j}}{dt}&=&F^{j}=-{\displaystyle\sum_{i,l=1}^nW^{ij}A_{li}v^l+\sum_{i=1}^nW^{ij}\frac{\partial L}{\partial q^{i}}}
\end{array}\right.
\end{equation}
which gives the second order differential equation
$$
\frac{d^{2} q^{j}}{dt^{2}}=F^{j}=-\sum_{i,l=1}^nW^{ij}A_{li}\frac{d q^{l}}{dt}+\sum_{i=1}^nW^{ij}\frac{\partial L}{\partial q^{i}}.
$$

In the case of  a  mechanical type Lagrangian system in the Riemanian manifold $(M,g)$, the Lagrangian is given by $L=T_{g}-V$ where 
$T_{g}$ is the kinetic energy defined by the metric and $V:M\to\mathbb{R}$ is the potential energy. The local coordinate expression for  the kinetic energy $T_g$ is
\begin{equation}
T_g(v)=\frac 12 \sum_{i,j=1}^n\,g_{ij}(\tau_M(v))\,v^iv^j\, .\label{kinenerg}
\end{equation}

To understand $V$ as a function defined in $TM$ we need to write $\tau_M^{*}V$. By simplicity we continue writing simply  $V$.

 For such kind of systems the Cartan 1-form $\theta_{T_g}$  given by      
 $\theta_{T_g}=d{L}\circ S=d{T_g}\circ S$,  reduces to 
$\theta_{L}(v)=\widehat g(v)=g(v,\cdot)$, which in local coordinates looks as 
\begin{equation}
\theta_{L}(q,v)=\sum_{i,j=1}^ng_{ij}(q)\,v^j\,dq^i, \label{cartan1}
\end{equation}
with associated  symplectic form $\omega_{L}$ given by 
\begin{equation}
\omega_{L}= \sum_{i,j=1}^ng_{ij}\, dq^i\wedge dv^j +\frac{1}{2}\sum_{i,j,k=1}^n\left(\frac{\partial g_{ij}}{\partial q^k}v^j-\frac{\partial g_{kj}}{\partial q^i}v^j\right) dq^i\wedge dq^k.\label{cartan2}
\end{equation}

 Here this 2-form $\omega_{L}$ is symplectic and consequently the Lagrangian $L=T_{g}-V$ is a regular Lagrangian as defined above. The matrix of 
$\omega_{L}$ is of  the form (\ref{omegaL}), 
but now with
$$
A_{ij}=\sum_{k=1}^n\left(\pd{g_{kj}}{q^i}- \pd{g_{ki}}{q^j}\right)v^k, \qquad 
W_{ij}=g_{ij}.
$$
 
 The total energy is  $E_L=T_{g}+V$, hence the dynamical vector field defined as in (\ref{dyneqL}) is given by (\ref{GammaL})   
with
\vspace{-4mm}
\begin{eqnarray}
F^i(q,v)&=&-\sum_{j,l=1}^nW^{ij}A_{lj}v^l+\sum_{j=1}^nW^{ij}\frac{\partial L}{\partial q^{j}}-\sum_{j=1}^ng^{ji}\frac{\partial V}{\partial q^j}\\
&=&-\sum_{j,k,l=1}^ng^{ij}\left(\pd{g_{kj}}{q^l}- \pd{g_{kl}}{q^j}\right)v^kv^l-\sum_{j=1}^ng^{ji}\frac{\partial V}{\partial q^j}\\
&=&-\sum_{k,l=1}^n\Gamma_{kl}^i v^kv^l-\sum_{j=1}^ng^{ji}\frac{\partial V}{\partial q^j}\, ,
\end{eqnarray}
where use has been made of  definition (\ref{Gijk}).

Once again the vector field $\Gamma_L$ corresponds to a second order differential equation we can write as
\begin{equation}\label{Gammasode}
\ddot q^i=-\sum_{k,l=1}^n\Gamma_{kl}^i \dot q^k\dot q^l-\sum_{j=1}^ng^{ji}\frac{\partial V}{\partial q^j}\, , \quad i=1,\ldots,n,
\end{equation} 
\noindent where the last term,  $X_V=-{\displaystyle\sum_{i,j=1}^ng^{ji} \frac{\partial V}{\partial q^i}\frac{\partial}{\partial v^j}}=-(\mathrm{grad}\; V)^v$, comes from the external force. Here $Y^v$ denotes the vertical lift of the vector field $Y\in \mathfrak{X}(M)$ (see e.g. \cite{Cr81,Cr83a}). We are in the usually called case of conservative systems or potential forces.

If $\mathrm{grad}\; V=0$,  the external force vanishes, we say we have a free motion, $L=T_g$, and then
 \begin{equation}\label{GammaTg}
\Gamma_{T_g}=\sum_{i=1}^n\left(v^i\pd{}{q^i}+F^i(q,v)\, \pd{}{v^i}\right)\, ,\quad F^i(q,v)=-\sum_{k,l=1}^n\Gamma_{kl}^i v^kv^l .
\end{equation}  
\subsection{General forces and Newtonian mechanical systems}

We can consider a more general situation where the external forces are not 
potential ones. This is the case when the force is given by a non-exact semibasic 1-form $F\in \bigwedge^1(TM)$; then the dynamics $\Gamma\in\mathfrak{X}(TM)$ is given by 
\begin{equation}
i(\Gamma)\omega_{T_g}-dE_{T_g}=F\, ,\label{generalF}
\end{equation}
instead of a free motion whose dynamics $\Gamma_{T_g}$ is given by $i(\Gamma_{T_g})\omega_{T_g}=dE_{T_g}$ as we know.

 The 1-form $F$ is called the {\sl force form} or work form and, being semibasic in $TM$, its local expression is $F={\displaystyle\sum_{i=1}^n}F_i(q,v)\,dq^i$. When  the 1-form $F$ is 
 basic, that is $F={\displaystyle\sum_{i=1}^n}F_i(q)dq^i$, it does not depend on the velocities, and then the system is said Newtonian. In this last case, given the force form $F$, then 
 $X_F={\displaystyle\sum_{i,j=1}^ng^{ij}F_i\frac{\partial}{\partial v^j}}$ is 
the local expression of the {\sl vector field of force} which is the vertical lift to $TM$ of the vector field in $M$, $Z_F=\widehat g^{-1}F$, i.e. $X_F=Z_F^v$. The above case of potential forces, 
with $L=T_g-V$, is given by $F=-\tau^*(dV)$ with $V:M\to\mathbb{R}$.
 
 Conversely, we can give a {\sl vector field of forces}, $Z\in\mathfrak{X}(M)$, with $Z={\displaystyle\sum_{i,j=1}^n}Z^i\partial/{\partial q^i}$, then $F_Z=i(Z)g={\displaystyle\sum_{i,j=1}^n}g_{ij}Z^idq^j$ is a 1-form in $M$ and $\tau^*F_Z$ is a basic 1-form in $TM$, the force form associated to the force vector field $Z$. In this approach the 
dynamical vector field $\Gamma$ is given by
 \begin{equation}
 i(\Gamma)\omega_{T_g}-dE_{T_g}=F_Z\, ,\label{generalF_Z}
 \end{equation}
  where we write $F_Z$ instead of $\tau^*F_Z$ for simplicity. The corresponding second order differential equation for its integral curves is
  \begin{equation}
  \ddot q^i=-\sum_{k,l=1}^n\Gamma_{kl}^i \dot q^k\dot q^l+Z^i\, ,\quad  i=1,\ldots,n.\label{newtonZ}
  \end{equation}
  
  Furthermore, we can consider the case where the vector field of forces is depending on the velocities $Z={\displaystyle\sum_{i=1}^n}Z^i(q,v){\partial}/{\partial q^i}$, that is, it is  a 
  vector field  along the projection $\tau_M:TM\to M$. In this case we proceed as above obtaining the 1-form of forces $F_Z=i(Z)g={\displaystyle\sum_{i,j=1}^n}g_{ij}Z^idq^j$ and thecorresponding expression for the dynamical vector field $\Gamma$.

 In all these situations the dynamical vector field $\Gamma$ is a second order differential equation with a similar local expression as above.
 
\section{Application to geodesics, free motion and geodesic fields}

\subsection{Curves with extremal length}
 As defined in (\ref{length}), the extremal length curves in $(M,g)$  are those corresponding to the action defined by the Lagrangian $\ell (v)= 
\sqrt{g(v,v)} $, even if we have to restrict ourselves to 
the open submanifold $T_0M=\{ v\in TM\mid v\ne 0\}$ in order to preserve the differentiability.  Remark that the length  does not depend on the parametrisation of the curve
and consequently a reparametrisation of an extremal  length curve leads to another extremal  length curve. 

Consider, therefore,  the Lagrangian
$\ell(v)=\sqrt{2\,{T_g}(v)}$,
where $T_g\in C^\infty(TM)$ is the function defined in (\ref{kinenerg})
\begin{equation}
T_g(v)=\frac{1}{2}\,g(v,v),\qquad v\in TM,\label{defTg}
\end{equation}  
usualy called {\it{kinetic energy}} in the context of mechanical systems.

 This function $\ell (v)\in C^\infty(T_0M)$
 is a singular Lagrangian that  has been studied in \cite{CL91}
 in a problem  with applications in a geometric approach to optics. This  
singular character of $\ell$ is responsible of the fact that a reparametrisation of an extremal  length curve gives rise to
 another  extremal  length curve.
 
 Being $\ell$ a singular Lagrangian, the symplectic approach is not  directly applicable but we can use the first method  and obtain the Euler-Lagrange equations. As $\ell(q, \dot q)=\sqrt {g_{ij}(q)\dot q^i\dot q^j}$, 
then
   $$
  \pd{\ell}{\dot q^i}=  \sum_{j=1}^n\frac{g_{ij}(q)\dot q^j} {\ell}, \qquad   \pd{\ell}{q^i}= \frac 1{2\ell} \sum_{j,k=1}^n\left(\pd {g_{jk}}{q^i}\dot q^j\,\dot q^k\right),\quad  i=1\ldots,n,
   $$
   and the Euler-Lagrange equations of the Lagrangian $\ell$ are:
   $$
   \frac d{dt}\left( \sum_{j=1}^n\frac{g_{ij}(q)\,\dot q^j} {\ell}\right)= \frac 1{2\ell}\left( \sum_{j=1}^n\pd {g_{jk}}{q^i}\dot q^j\,\dot q^k\right),\quad  i=1\ldots,n,
   $$
   where  $t$ is the parameter of the solution curves.
   
   If we parametrise the curves by the arc-lenght $s$,  that is we apply a Sundman transformation, then as $ds/dt=\ell$, $ \dot q^i=\ell\, q^{\prime i} =\ell \,dq^i/ds$, we have that 
   the previous system becomes 
   \begin{equation}
   \frac d{ds}\left(\sum_{j=1}^ng_{ij}(q)\,q^{\prime j}\right)=\frac 12 \sum_{j,k=1}^n\pd {g_{jk}}{q^i}\,q^{\prime j}\,q^{\prime k},\label{geodesicss}
      \end{equation}
      i.e. 
    \begin{equation}
   q^{\prime\prime i}=\sum_{j,k,l=1}^ng^{il}(q)\left(\frac 12 \pd {g_{jk}}{q^l}-\pd{g_{lj}}{q^k}\right)\,q^{\prime j}\,q^{\prime k},\label{geodesicss2}
      \end{equation}
      or in another form, 
   \begin{equation}
 q^{\prime\prime i}+\sum_{j,k=1}^n\Gamma^i_{jk}(q)\, q^{\prime j}\,q^{\prime k}=0,\label{eqgeod}
\end{equation} 
where  $\Gamma_{jk}^i$ are the Christoffel symbols of the second kind   defined by the metric $g$.
 
Equation (\ref{eqgeod}), which is the same equation as (\ref{geodeq}), is called {\it geodesics' equation} obtained from 
Euler-Lagrange equations for $\ell\,$ by the above Sundman transformation.
 It is to be remarked that the same can be said for a Sundman reparametrisation $d\tau /dt=a\, \ell$, where $a\in \mathbb{R}$. This corresponds to the fact that the parameter of geodesic curves is defined up to an affine transformation $s\mapsto a\, s+b$. See Subsection 5.3 for a reciprocal of this result and some comments on it.

\subsection{Free motions on a Riemann manifold}  

   Free motion on a Riemann manifold $(M,g)$  is described by a regular Lagrangian given by the function $T_g\in C^\infty(TM)$, as defined in (\ref{kinenerg}),
which  is the kinetic energy defined by the metric. 

 We can follow the geometric approach with the Lagrangian $L=T_g$, which reduces to the kinetic energy and is a regular Lagrangian. The Cartan 1-form $\theta_{T_g}$  and the symplectic 2-form are given by
 \begin{equation}
  \begin{array}{rcl}
   \theta_{T_g}(q,v)&=&{\displaystyle \sum_{i,j=1}^n}g_{ij}(q)\,v^j\,dq^i,\\
 \omega_{T_g}&=&{\displaystyle \sum_{i,j=1}^n} g_{ij}\, dq^i\wedge dv^j +{\displaystyle\frac{1}{2} \sum_{i,j,k=1}^n\left(\frac{\partial g_{ij}}{\partial q^k}v^j-\frac{\partial g_{kj}}{\partial q^i}v^j\right) }
dq^i\wedge dq^k.
\end{array}
 \end{equation}
 The total energy is $E_{T_g}=T_g$ and the dynamical vector field solution of $i(\Gamma_{T_g}) \omega_{T_g}=dE_{T_g} $ is
 \begin{equation}
\Gamma_{T_g}= \sum_{i=1}^n\left(v^i\pd{}{q^i}+F^i(q,v)\, \pd{}{v^i}\right), \quad F^i(q,v)=-\sum_{j,k=1}^n\Gamma^i_{jk}(q)v^jv^k,\label{GTg}
\end{equation}
where, as above,   $\Gamma_{jk}^i$ denote  the second class Christoffel symbols defined by the metric.

The projected curves on the base manifold of the integral curves of the  vector field $\Gamma_{T_g}$ are solutions of the system of 
second order differential equations 
\begin{equation}
\ddot q^i+\sum_{j,k=1}^n\Gamma^i_{jk}(q)\,\dot q^j\,\dot q^k=0,\label{ELeqsg} 
\end{equation}
which is the local expression for  
$\nabla_{\dot \gamma}\,\dot \gamma=0$, where $\gamma$ is a curve in $M$ 
 and $\dot \gamma$ the vector field 
along $\gamma$ given by the velocities of the curve at each point.
This shows that the above integral curves are geodesics of the Riemannian 
structure, and  therefore, the geodesic curves $\gamma$ are such that their tangent prolongations $\dot\gamma$  
are integral curves of $\Gamma_{T_g}$.

  In particular, as indicated above,  an  extremal length curve parametrised by its arc-length 
is a solution of
   \begin{equation} 
 \frac{d^2q^i}{ds^2}+\sum_{j,k=1}^n\Gamma^i_{jk}\, \frac{dq^j}{ds}\, \frac{dq^k}{ds}=0,\label{geodeqs}
\end{equation}
and therefore in terms of such parametrisations the extremal length curves are geodesics of the corresponding metric.

As we will see in Subsection 5.3, if a curve $\gamma(t)$ satisfies $\nabla_{\dot\gamma}\dot\gamma=0$, then $g(\dot\gamma,\dot\gamma)$ is constant along $\gamma$, hence the parameter $t$ is a  real affine  function of the arc-length, that is $t=as+b$ with $a,b\in\mathbb{R}$.

\subsection{Geodesic fields and Sundman transformation}

There are distinguished classes of vector fields on  a Riemann manifold $(M,g)$. For instance,  {\it Killing vector fields }$X$ are those such that $\mathcal{L}_Xg=0$, 
i.e. vector fields whose   local flows are preserving $g$.  Another relevant class is that of autoparallel vector fields, characterised by the property $\nabla_XX=0$, where $\nabla$ is the corresponding Levi--Civita connection.

 Remark that $X \in  \mathfrak{X}(M)$ is an  autoparallel vector field  if and only if   its integral curves $\gamma$ are geodesics, i.e.  they satisfy the equation
$\nabla_{\dot \gamma}\,\dot \gamma=0$, and then when  lifted to the
tangent bundle $TM$ are integral curves of $\Gamma_{T_g}$, the second order geodesic vector field given in (\ref{2ogeodvf}) or  (\ref{GTg}). In fact, suppose that $\nabla_XX = 0$ and 
let $\gamma : I \subset  \mathbb{R} \to  M$ be an integral curve of $X$, that is
$\dot \gamma = X  \circ  \gamma$. Then:
$$ \nabla _{\dot \gamma (t)}\dot \gamma(t) =  \nabla _{(X \circ \gamma)(t)}(X  \circ  \gamma)(t) = ( \nabla_XX)(\gamma(t)) = 0 ,$$
and therefore $\gamma$ is a geodesic curve in $M$. If, on the contrary, we suppose that every integral curve of $X$ satisfies the equation 
  $ \nabla _{\dot \gamma (t)}\dot \gamma(t) = 0$,  let
$p \in  M$  and $\gamma : I \subset  \mathbb{R} \to  M$ be the integral curve of $X$ with initial condition $p$, we have
$$( \nabla_XX)(p) =  \nabla_{ X(p)}X =  \nabla_{\dot\gamma (0)}\dot \gamma (t) = ( \nabla_{\dot \gamma (t)}\dot \gamma (t))_{|t=0} = 0 .$$
But as this is true  for every $p \in  M$, we have $ \nabla_XX = 0$. 

Because of this property the  autoparallel 
vector fields are also called {\it geodesic vector fields} \cite{DPBT}.   It is to be remarked that for each real number $a$, if  $X$ is a geodesic vector field, then $a\, X$ is geodesic too.

There exists an even  more general class of 
{\it generalised geodesic vector fields}, usually  also called {\it pregeodesic vector fields}, which are those such that there exists a  function $f$ satisfying $\nabla_XX=f\, X$. 

Before going to the application of Sundman transform to these vector fields, we recall some known results with slightly new proofs.

In  \cite{BN09},  it is proven,  Lemma 3, and in \cite{N19} is used, that a Killing vector field $X \in \mathfrak{X}(M)$   has constant length
if and only if every integral curve of $X$ is a geodesic, that is,  $X$  is a geodesic vector field. We can give a little more general proof:

a) Using that the connection $\nabla$ is metric and torsionless,   we have on the one  hand that for arbitrary vector fields $X,Y\in \mathfrak{X}(M)$,
$$
\begin{array}{rcl}
\mathcal{L}_X(g(X,Y))&=&(\mathcal{L}_Xg)(X,Y)+g(\mathcal{L}_XX,Y)+g(X,\mathcal{L}_XY)\\&&\\
&=&(\mathcal{L}_Xg)(X,Y)+g(X,\nabla_XY)-g(X,\nabla_YX)\\&&\\
&=&(\mathcal{L}_Xg)(X,Y)+g(X,\nabla_XY)-{\displaystyle\frac12}\mathcal{L}_Y(g(X,X)),
\end{array}
$$
and on the other hand, 
\begin{equation}
\mathcal{L}_X(g(X,Y))=g(\nabla_XX,Y)+g(X,\nabla_XY)\, .\label{metricXXY-2}
\end{equation}
Comparing both expressions we have that
$$
g(\nabla_XX,Y)=(\mathcal{L}_Xg)(X,Y)-\frac 12\mathcal{L}_Y(g(X,X)).
$$

b) If $X=Y$, the  expression (\ref{metricXXY-2}) reduces to  $\mathcal{L}_X(g(X,X))=2g(\nabla_XX,X)$ and this relation shows that if  $X$ is a geodesic vector field, $\nabla_XX=0$,
then 
$\mathcal{L}_X(g(X,X))=0$, that is, the norm of $X$ is constant along the integral curves of $X$.

c) Moreover, if $X$ is a Killing vector field, we have that $g(\nabla_XX,Y)=-(1/2)\mathcal{L}_Y(g(X,X))$,  and, consequently, as  $Y$ is an arbitrary vector field, we see that 
  then  $X$ is a geodesic vector field, i.e. $\nabla_XX=0$,  if and only if $g(X,X)$ is a real constant as we wanted. 
  
   \bigskip
A remarkable similar property was proved in \cite{DCh18}: a generalised geodesic 
vector field $X$ of constant length on a
Riemannian manifold  is a geodesic vector field, because
   if $\varphi$ is the function defined by $\varphi=\frac 12g(X,X)$, as  $\nabla_XX=f\, X$, we have that  
$$\mathcal{L}_X(\varphi)=\frac 12g(\nabla_XX,X)+\frac 12g(X, \nabla_XX)=2 f\varphi,$$ 
and equivalently,  $2f= \mathcal{L}_X(\log \varphi )$, 
which shows that  as $\varphi$ is constant,  then $f=0$, and therefore $X$ is a geodesic vector field.

As a time-reparametrisation of  an extremal  length curve gives rise to
 another  extremal  length curve, we can study the reparametrisation of geodesics curves, which we know are parametrised by the arc-length.
If we consider a reparametrisation of a solution of (\ref{geodeqs}) with a 
generic parameter $\tau$ defined, as a given function of $s$, by
    \begin{equation}
\frac{ds}{d\tau}=\xi(\tau),\label{repar}
    \end{equation}
we have 
   \begin{equation}
   \frac {d}{ds}=\frac 1\xi \frac {d}{d\tau}\Longrightarrow \frac{d^2}{ds^2}=\frac 1\xi \frac {d}{d\tau}\frac 1\xi \frac {d}{d\tau}=-\frac{d\xi/d\tau}{\xi^3} \frac {d}{d\tau}+\frac 1{\xi^2} \frac{d^2}{d\tau^2},  \label{d2s}  
\end{equation}
and the corresponding differential equation  (\ref{geodeqs})  for geodesics in such a generic parametrisation is
    \begin{equation} 
 \frac{d^2q^i}{d\tau^2}+\sum_{j,k=1}^n\Gamma^i_{jk}\, \frac{dq^j}{d\tau}\, \frac{dq^k}{d\tau}= \lambda (\tau) \frac {dq^i}{d\tau} ,\label{geodeqtau}
\end{equation}  
with
  \begin{equation}
\lambda(\tau)=  \frac{d^2s}{d\tau^2}\left(\frac{ds}{d\tau}\right)^{-1},\label{lambda}
  \end{equation}
that can also be written as 
$$\lambda(s)= - \frac{d^2\tau}{ds^2}\left(\frac{ds}{d\tau}\right)^{2}.
$$
Note that when $\xi(\tau)$ is a constant $a$, that corresponds to an affine change of parameter, we obtain from (\ref{lambda}) that  $\lambda=0$, i.e. (\ref{geodeqtau}) reduces
 to the original equation. This corresponds to the case $\tau=as+b$, wth $a,b\in\mathbb{R}$, that is,
  ${\displaystyle \frac{d^2\tau}{ds^2}=0}$, as we commented at the end of Subsection 5.1.

 Conversely, a solution of the differential equation (\ref{geodeqtau})  corresponds to a solution of (\ref{geodeqs}) 
with a reparametrisation defined by (\ref{repar}) with $\xi(\tau)=\exp\left({\displaystyle \int^\tau}\lambda(\zeta)\, d\zeta\right)$.

 Instead of a reparametrisation (\ref{repar}) we can consider  a classical Sundman transformation 
\begin{equation}
\frac{ds}{d\tau}=f(q),\label{repar1}
    \end{equation}
    and then differential equation (\ref{geodeqs}) for geodesics becomes
       \begin{equation} 
 \frac{d^2q^i}{d\tau^2}+\sum_{j,k=1}^n\Gamma^i_{jk}\, \frac{dq^j}{d\tau}\, \frac{dq^k}{d\tau}=\frac 1f \left(\sum_{k=1}^n \pd f{q^k} \frac {dq^k}{d\tau}\right) \frac {dq^i}{d\tau} =\frac d{d\tau}(\log f)\,  \frac {dq^i}{d\tau}.\label{geodeqtau2}
\end{equation}

Now we try to obtain the relation between geodesic fields and Sundman transformation in a more geometric way.
 
 If we carry out a  Sundman transformation   (\ref{genSundman}) in the description of the geodesics of a Riemann manifold $(M,g)$ as integral curves of geodesic vector fields, a geodesic vector field $X$ should be 
 replaced by the vector field $Y=f\, X$, and then its integral curves are not geodesics because their  velocities are not covariantly constant. 
 Actually, from 
 $$\nabla_{f\,X}(f\,X) =f\,\nabla_{X}(f\,X) = f^2 \,\nabla_XX +f\,(Xf) X,$$ 
 we see that $\nabla_XX =0$ implies that $\nabla_{f\,X}(f\,X) =(Xf)\,(f\,X)$,
 i.e. $\nabla_YY =(Xf)Y$. Conversely, 
the integral curves of a vector field $X \in  \mathfrak{X}(M)$ can be reparametrised to be
geodesic curves if and only if there exists a function $f : M  \to  \mathbb{R}, \, f > 0$, such that  $\nabla_XX = f\, X$, 
because in this case, for any function   $\lambda  : M  \to  \mathbb{R},\,  \lambda  > 0$,  we have
$$ \nabla _{ \lambda X}( \lambda X) =
 ( \lambda  \mathcal{L}_X \lambda )X +  \lambda^2 \nabla XX = ( \lambda 
 \mathcal{L}_X \lambda )X +  \lambda^2fX =
  ( \lambda\,   X (\lambda)  +  \lambda^2f )X ,
 $$
 and taking as $ \lambda$  any solution of the differential equation $ \mathcal{L}_X (\log  \lambda)  = -f$, we have that  $\nabla_{  \lambda X}( \lambda\, X) = 0$,
and then  $\lambda X$ is a geodesic field in $M$ and its integral curves are
geodesic ones.

Observe that if  $\nabla_XX = fX$, i.e. $X$ is a generalised geodesic vector field, then $Y =  \lambda X$ with  $\mathcal{L}_X (\log  \lambda) 
 = -f$ is a geodesic vector field.

Summarising, we have proved:

\begin{prop}\label{geodrepar}
The integral curves of a vector field $X\in\mathfrak{X}(M)$ can be transformed by a Sundman transformation	 to be geodesic curves if, and only if, the vector field $X$ satisfies $\nabla_XX=fX$ for some nonvanishing  function $f:M\to\R$.
\end{prop}

\section{Applications in mechanical systems}
 
\subsection{Newtonian systems and time reparametrisation}

In this section we study  a mechanical Newtonian system determined by a vector field of forces $Z\in\mathfrak{X}(M)$ in the Riemannian manifold $(M,g)$. We know from (\ref{generalF_Z}) that the  corresponding dynamical vector field $\Gamma$ satisfies the equation
\begin{equation}
i(\Gamma)\omega_{T_g}-dE_{T_g}=F_Z\, ,
\end{equation}
where $F_Z=i(Z)g$. The integral curves of $\Gamma$ satisfy the second order differential equation
\begin{equation}
\ddot q^i=-\sum_{k,l=1}^n\Gamma_{kl}^i \dot q^k\dot q^l+Z^i\, ,\quad  i=1,\ldots,n.
\end{equation}
But a curve $\gamma:I\subset\mathbb{R}\to M$ satisfies this equation if and only if it satisfies the following one
\begin{equation}
\nabla_{\dot\gamma}\dot\gamma=Z\circ\gamma,
\end{equation}
as we can prove expressing both equations in coordinates.

We are now interested in studying the properties of vector fields $X\in\mathfrak{X}(M)$ playing the r\^ole of the geodesic vector fields  but in presence of a non-null 
vector field of forces $Z\in\mathfrak{X}(M)$.

The main result, to be compared with a previous one for the  case $Z=0$, is the following:

\begin{prop}
 Given $Z\in\mathfrak{X}(M)$,  the integral curves $\gamma$ of $X \in  \mathfrak{X}(M)$ satisfy the equation $\nabla_{\dot\gamma}{\dot\gamma}= Z 
\circ {\gamma}$,  if and only
if the vector field $X$ satisfies the equation $\nabla_XX = Z$.
\end{prop}
{\sl Proof.-}  Suppose that
  the integral curves $\gamma$ of $X \in  \mathfrak{X}(M)$ satisfy the equation $\nabla_{\dot\gamma}{\dot\gamma}= Z \circ  {\gamma}$.
If  $p$ is  a point of $M$   and $\gamma$  is the integral curve of $X$  starting from $p$, we have:
$$(\nabla_XX)(p) = \nabla_{X(p)}X = \nabla_{\dot \gamma(0)}(X \circ \gamma)(t) = \nabla_{\dot \gamma(0)}{\dot \gamma(0)}(t) = Z(\gamma(0)) 
= Z(p) ,
$$
and as  this is valid for any $p\in  M$, then $ \nabla_XX = Z$. 

Conversely, if $\nabla_XX =Z$  and $\gamma$  is an integral curve of $X$, that is $\dot \gamma = X \circ \gamma$, 
then $(\nabla_XX)( \gamma(t) )= Z( \gamma(t))$, which gives $\nabla_{\dot  \gamma(t)}\dot  \gamma (t) = Z(\gamma(t))$.   

 $\hfill\Box$
 
 \bigskip
 This result can be particularized to the case $Z=-\textrm{grad\,} V$ corresponding to a simple mechanical type system.

\bigskip
The relation with Sundman transformation is the following:

Suppose that the vector field $X \in   \mathfrak{X}(M)$ satisfies the equation $\nabla_XX = Z$ and consider
a Sundman reparametrisation of its integral curves, that is,  consider the vector field $Y = h\,X$, with $h$ a function,
$h : M \to  \mathbb{R}, \, h > 0$. Then   the corresponding equation for  
the vector field
$Y = h\,X$ is:
$$\nabla_Y Y = \nabla_{h\,X}(h\,X) = h\,(\mathcal{L}_Xh)X + h^2\nabla_XX = (\mathcal{L}_Y (\log h))Y + h^2\,Z = \lambda Y + h^2Z , 
$$
where $\lambda  = \mathcal{L}_Y (\log h) ={\displaystyle \frac 1 h} \mathcal{L}_Y h$.

This last equation, $\nabla_Y Y = \lambda Y + h^2\,Z$, is the general expression of time reparametrisation of the equation  $\nabla_XX = Z$, when 
changing the vector field $X$ by $Y = h\,X $ in the Newtonian system $\nabla_XX=Z$. 

This result can be summarised  as:

\begin{prop}
{\it The integral curves of a given vector field $Y \in   \mathfrak{X}(M)$ can be reparametrised to be
solutions of the Newtonian equation  $\nabla_{\dot  \gamma(t)}\dot  \gamma (t) = Z (\gamma(t))$  if, and only if, there exists a function 
$h : M \to  \mathbb{R}, \, h > 0$, which satisfies the equation $$\nabla_Y Y = \lambda Y + h^2Z,
$$
where $\lambda  = \mathcal{L}_Y (\log h) ={\displaystyle\frac 1 h} \mathcal{L}_Y h$}. $\hfill\Box$
\end{prop}

Remark that, as a consequence of this last Proposition, unless $Z = 0$  
or $ Z = f\, X$  with $\nabla_XX = 0$, we cannot find a Sundman reparametrisation of the curves solution to
$\nabla_{\dot  \gamma(t)}\dot  \gamma (t) = Z (\gamma(t))$ to obtain geodesics of the metric $g$, that is curves satisfying equation $\nabla_{\dot  \gamma(t)}\dot  \gamma (t) = 0$, and therefore, in order  to understand such curves
  as geodesics we need to change the Riemannian metric $g$ to eliminate
the external force with an adequate connection and, as we will see in the 
sequel,
 we cannot do this in the whole configuration
manifold $M$. Later on we will show how to proceed in the case of a mechanical type Lagrangian system,  leading to the so called  Jacobi
metric.
 
\subsection{Mechanical type systems on a Riemann manifold}  \label{mtsRm}

We consider now Lagrangian systems of simple mechanical type, that is,  Lagrangians of the form $L=T_g-\tau_M^*V$, defined by a Riemann metric $g$ on $M$ and  a function $V:M\to\mathbb{R}$. Its energy function is then $E=T_g+\tau_M^*V$. We have shown in (\ref{Gammasode}) that the second order differential equation of the dynamical trajectories is 
\begin{equation}
\ddot q^i+\sum_{j,k=1}^n\Gamma^i_{jk}(q)\,\dot q^j\,\dot q^k=-\sum_{l=1}^ng^{il}\pd V{q^l}, \quad 
i=1,\ldots,\dim M.\label{dyneq2}
\end{equation}

Which equation is the corresponding one  when we apply a time reparametrisation?

When expressed in terms of a 
generic parameter $\tau$ defined as a given function of $t$, as in (\ref{repar}), 
    \begin{equation}
\frac{dt}{d\tau}=\xi(\tau),\label{repar2}
    \end{equation}
we obtain as in (\ref{d2s}) that
    \begin{equation}
\frac {d}{dt}=\frac 1\xi \frac {d}{d\tau}\Longrightarrow \frac{d^2}{dt^2}=\frac 1\xi \frac {d}{d\tau}\frac 1\xi \frac {d}{d\tau}=-\frac{d\xi/d\tau}{\xi^3} \frac {d}{d\tau}+\frac 1{\xi^2} \frac{d^2}{d\tau^2}, \label{dtdtau}
    \end{equation}
and the differential equation for those curves in such a generic parametrisation will be 
    \begin{equation} 
 \frac{d^2q^i}{d\tau^2}+\sum_{j,k=1}^n\Gamma^i_{jk}\, \frac{dq^j}{d\tau}\, \frac{dq^k}{d\tau}- \lambda (\tau) \frac {dq^i}{d\tau} = -\xi^2(\tau)\sum_{l=1}^n\,g^{il}\pd V{q^l}, \quad i=1,\ldots,\dim M,\label{geodeqtau3}
\end{equation}  
with
  \begin{equation}
\lambda(\tau)= \frac 1{\xi} \frac{d\xi}{d\tau}.\label{lambda2}
  \end{equation}   
In this way we have obtained the differential equation of the trajectories of the given mechanical system in a general parametrisation.

\section{Conformal metrics and Sundman transformation. The Jacobi metric}

  \subsection{Sundman transformation in free motions. Changes in the Riemannian metric}
  
 In this Subsection and the next one we go to a more geometric approach to the same problem of reparametrisation. We will try to change the metric $g$ in order to modify the geometric properties of the dynamical trajectories of the system.
    
Let us remark that if we carry out the Sundman transformation   (\ref{genSundman})   in the framework  of Lagrangian formulation, 
with $f$ being a basic function,   the new velocities $\bar v$ are related to the previous ones $v$ by $\bar v=f\, v$,  and, moreover,
  if a system was defined by a Lagrangian $L$, in  order to the action be 
well defined and preserved,   the system must be described in terms of the new time $\tau$ by a new Lagrangian 
$$
\bar L(q,\bar v)=f\, L\left(q,\frac{\bar v} f\right), 
$$
 and then ${\displaystyle \int L\, dt= \int \bar L\, d\tau}$.
 
 The new dynamical vector field will be given by
 $$
  \Gamma_{\bar L}(q,\bar v)= \sum_{i=1}^n\left(\bar v^i\pd{}{q^i}+\bar F^i(q,\bar v)\, \pd{}{{\bar v}^i}\right),
  $$
 for appropriate functions   $\bar F^i$ as given by (\ref{GammaLcoor}). 
 
 In the particular case of a free motion on a Riemann manifold $(M,g)$, the new Lagrangian will be the  kinetic energy defined by the metric 
 $(1/f)\,g$, because $g$ is quadratic in velocities and therefore $f\, L(q,\bar v/f)=(1/f) L(q,\bar v)$.
 
 On the other side,  the considered
Sundman transformation for a mechanical type system amounts to change not only the Riemann structure
from $g$ to the conformally related one  $\bar g=(1/f) g$ but also the potential function $V$ to
$\bar V=f\, V$, a  fact that has recently been used in the corresponding Hamiltonian formalism  in the search for a new superintegrable system \cite{CRS} according to the metod developed in \cite{Liouv49}. Recall that  a mechanical type system for which there is a coordinate system such that  the potential function is a sum 
$V(q)=V_1(q_1)+\ldots+V_n(q_n)$  and the local expression of the Riemann structure is diagonal, i.e.
  \begin{equation}
 L(q,v)=\frac 12 \sum _{i=1}^n a_i(q_i) v_i^2-  \sum _{i=1}^n V_i(q_i),\label{Vdescomp}
  \end{equation}   
is separable as a sum of one-dimensional systems  and hence integrable by quadratures. 
A generalisation of such system is due to Liouville  \cite{Liouv49} and consists on the Hamiltonian 
  \begin{equation}
 H(q,v)=\frac 1{2W(q)} \sum _{i=1}^n a_i(q_i) v_i^2+  \frac 1{W(q)}\sum _{i=1}^n V_i(q_i),
 \label{HLiouv}
  \end{equation}   
  where $W(q)=W_1(q_1)+\ldots+W_n(q_n)$. 
These systems are called Liouville systems \cite{Perelomov,GzLMateos19} and one can check that the $n$ functions 
$$ 
 F_i=\frac 12 a_i(q)\, p_i^2+V_i(q)-W_i\,H,  \quad i=1,\ldots,n,
$$
are constants of motion $\{H, F_i\}=0$, but they are not independent because $\sum_{i=1}^n F_i=0$.
Therefore, starting from an appropriate Hamiltonian $H$ one can look for possible functions  $f$ such that the new Hamiltonian $f H$ satisfies the required properties of superintegrability. 

This method of deriving from a given set of involutive functions a new one  recalls very much the St\"ackel transforms \cite{BKM} and the coupling constant metamorphosis 
 \cite{HGDR} where also infinitesimal time reparametrisations are used (see also \cite{SB08} where reciprocal transformations of different times are used in the study of generalised 
 St\"ackel transforms).

 It is interesting to study the relationship with the case of a 
metric $\bar g=e^{2\varphi}g$, which corresponds to $f=e^{-2\varphi}$. 
 We have proven in (\ref{bGijk}) that in this case  the new   Christoffel 
symbols of the second kind are:
 \begin{equation}
\bar \Gamma^i_{jk}(q)= \Gamma^i_{jk}(q)+\delta^i_j\pd{\varphi}{q^k}+\delta^i_k\pd{\varphi}{q^j}-\sum_{l=1}^ng_{jk }\,g^{il}\pd{\varphi}{q^l}, \quad i,j,k=1,\ldots,\dim M. %\label{bGijk} 
\end{equation}

Then, instead of (\ref{ELeqsg}) the  new equations of the motion for a mechanical system defined by $\bar g$ and $V$ will be:
 \begin{equation}
 \frac {d^2q^i}{d\tau^2}+\sum_{j,k=1}^n \Gamma^i_{jk}\frac{dq^j}{d\tau}\frac{dq^k}{d\tau}+2\sum_{k=1}^n \pd{\varphi}{q^k}\frac {dq^k}{d\tau}\frac {dq^i}{d\tau}-\sum_{j,k,l=1}^n g_{jk }\frac{dq^j}{d\tau}\frac{dq^k}{d\tau}\,g^{il}
\pd{\varphi}{q^l}=-\sum_{j}^n e^{-2\varphi} g^{ji}\frac{\partial V}{\partial q^j}, \label{fieqs}
 \end{equation} 
  for $ i=1,\ldots,\dim M$.

Here we see that the two last terms
 of the left-hand side of (\ref{fieqs}) correspond to the two last terms in (\ref{bGijk}).

If we reparametrise  in (\ref{fieqs}) in a way similar to (\ref{repar2}), 
i.e.  
 \begin{equation}
 \frac {d\tau}{d\eta}=\xi(\eta),\label{repar3}
  \end{equation}
using expressions corresponding to (\ref{dtdtau}) such differential equation becomes
 \begin{equation}
\frac {d^2q^i}{d\eta^2}+ \sum_{j,k=1}^n \Gamma^i_{jk}\frac{dq^j}{d\eta}\frac{dq^k}{d\eta}-\lambda(\eta)\frac{dq^i}{d\eta}+2\sum_{k=1}^n\pd{\varphi}{q^k}\frac {dq^k}{d\eta}\frac {dq^i}{d\eta}-\sum_{j,k,l=1}^ng_{jk }\frac{dq^j}{d\eta}\frac{dq^k}{d\eta}\,g^{il}\pd{\varphi}{q^l}=
-\xi(\eta)^2\sum_{j=1}^ne^{-2\varphi}g^{ji}\frac{\partial V}{\partial q^j},\,\, \label{fieqs2}
 \end{equation}
 for $ i=1,\ldots,\dim M$,  where 
 \begin{equation}
\lambda(\eta)= \frac 1{\xi} \frac{d\xi}{d\eta}.\label{lambda3}
  \end{equation}
  The expression (\ref{fieqs2}) is the differential equation for the dynamical trajectories of a simple mechanical control system in a general parametrisation with respect to a metric $\bar g=e^{2\varphi}g$.
  
  It is to be remarked that in the free (geodesic) case $V=0$ the equation of geodesic curves for the Riemann structure  $\bar g=e^{2\varphi}g$ in terms of the new time $\tau$ such that $ d\tau= e^{2\varphi} dt$ is
 \begin{equation} 
  \frac {d^2q^i}{d\tau^2}+\sum_{j,k=1}^n \Gamma^i_{jk}\frac{dq^j}{d\tau}\frac{dq^k}{d\tau}+2\sum_{k=1}^n \pd{\varphi}{q^k}\frac {dq^k}{d\tau}\frac {dq^i}{d\tau}-\sum_{j,k,l=1}^n g_{jk }\frac{dq^j}{d\tau}\frac{dq^k}{d\tau}\,g^{il}
\pd{\varphi}{q^l}= 0, \label{figeodeqs}
 \end{equation} 
 which does   not  coincide with (\ref{geodeqtau2}) for $f=e^{-2\varphi}$  because of the last term on the leftt-hand side of (\ref{figeodeqs}).

 \subsection{Sundman transformation and the Jacobi metric}  
  From a local viewpoint we can look for a possible choice of the function $\varphi$  such that there exists a Sundman reparametrisation (\ref{repar1})  of the geodesics lines equation 
  in $(M,g)$  given by (\ref{eqgeod}) leading  to  the   equation  (\ref{fieqs}). 
 This choice is possible when  we restrict our study to motions given by (\ref{dyneq2}) with a fixed energy $E$, i.e. when 
   \begin{equation}
   \sum_{j,k=1}^ng_{jk }\frac{dq^j}{d\eta}\frac{dq^k}{d\eta}=f^2 . \label{fixE}
\end{equation}
  
  A simple comparison of (\ref{geodeqtau2}) with (\ref{fieqs}) shows that both equations coincide when
     \begin{equation}
  \frac d{dt}(\log f)=-2\frac {d\varphi}{d\tau} ,\label{cond1}
     \end{equation}
   i.e. $f=e^{-2\varphi}$,  and
      \begin{equation}
   f^2 \pd{\varphi}{q^i}= e^{-2\varphi}\pd V{q^i}.\label{cond2}
     \end{equation}

   When the energy $E$  of the mechanical system defined by $\bar g$ and $V$ is fixed, 
   $$
   e^{2\varphi}\sum_{j,k=1}^n g_{jk}\pd{q^j}{\tau}\pd{q^k}{\tau}=2(E-V),
   $$
   from where we obtain
   $f^2=2\, e^{-2\varphi} \,(E-V)$. Using this value for $f^2$,  (\ref{cond2}) becomes 
   $$2(E-V)\pd{\varphi}{q^i}=\pd V{q^i},
   $$
and from here we obtain that 
\begin{equation}
\varphi(q)=-\frac12 \log (E-V) \Longrightarrow e^{2\varphi} =  (E-V)^{-1}. 
\end{equation}
 
Consequently, we have recovered the well-known result about Jacobi metric: The geodesics of the Jacobi metric $g_E=(E-V)g $ when using as parameter 
its arc-length $s_E$,
are the solutions with fixed energy $E$ of the mechanical system determined by the Lagrangian $L_{g,V}$.

\bigskip

With a simple look at  equations (\ref{geodeqs}) and (\ref{dyneq2}) or the corresponding equations in terms of $\nabla_X $,  one can try to relate 
both types of equations.
From the geometric viewpoint this suggests us  the  following question: 
Can we conformally change the Riemannian metric $g$ to another one $\bar g=\exp(2\varphi)\, g$, with an adequate $\varphi$, such that the vector 
fields $X$ solution to the dynamical equation,  $\nabla_XX=-\textrm{grad\,} V$, are geodesic, or at least pregeodesic,   vector fields  for the Riemannian metric $\bar g$?

We have the following result:

\begin{theorem}\label{HJ-metric}
Let $X\in\mathfrak{X}(M)$ be a solution to the equation  $\nabla_XX=-{\rm grad\,} V$, and suppose that the function $E(X):M\to \mathbb{R}$ defined  on $M$ by  $E(X)=E\circ X$, i.e. $E(X)= (T_g+V\circ\tau_M)(X)$, takes a constant value, $E=E_0\in\mathbb{R}$.
Then, $X$ is a pregeodesic vector  field for the Riemannian metric  $\bar 
g=\exp(2\varphi)\, g$ if, and only if, there exists a function $f$  
such that
$$2(E_0-V)\,{\rm grad}_g\,\left(\varphi- \log(E_0-V)^{1/2}\right)=-f\, X.$$

Moreover, if $f=0$ and $M$ is connected, then 
 $\varphi= \log(E_0-V)^{1/2}+k$,  where $k$ is a real number,  that is, $\bar g=e^{2k}(E_0-V)g$.
The case $k=0$ is usually called Jacobi metric: $\bar g=(E_0-V)g$.
\end{theorem}

{\sl Proof.-} Let be $\bar g=\exp(2\varphi)\, g$, and recall the relation (\ref{barnabla1}) between the covariant derivatives with respect to $g$ and $\bar g$  and the corresponding equation
(\ref{barnabla3}) for $X=Y$. 

Therefore,  the vector field $X$  is pregeodesic for $\bar g$  if, and only if, there exists a function $h$  such that  $\bar \nabla_XX=h\, X$, and hence if, and only if,
$\nabla_XX-g(X,X)\, {\textrm{grad}}_g\varphi=f\, X$ with $f=h-2 \mathcal{L}_X\varphi$.

But  the condition $\nabla_XX=-{\textrm{grad}_g}V$ for the vector field $X$  is then equivalent to
$$
2(E_0-V)\,{\textrm{grad}}_g\left(\varphi- \log(E_0-V)^{1/2}\right)=-f\, X, 
$$
where use has been made of $E(X)=E_0$ and 
$$
\textrm{grad}_g\left(\log(E_0-V)^{1/2}\right)=-\frac{1}{2(E_0-V)}\,\textrm{grad}_g V.
$$

If $f=0$, then ${\textrm{grad}}_g\left(\varphi- \log(E_0-V)^{1/2}\right)=0$, and consequently,  when $M$ is connected, there exists a real number $k$ such that 
 $\varphi=\log(E_0-V)^{1/2}+k$, where $k$ is a real number.

We have obtained that  $\bar g= \exp(2\varphi)g= e^{2k}(E_0-V)g$. Obviously the factor $k$ is irrelevant for the characterization of the geodesics of the family of metrics $\bar g$ and we can choose $k=0$ as in the classical Jacobi metric.

 $\hfill\Box$
 
 The above Theorem is, at our knowledge, the most natural approximation to the description of the Jacobi metric in a mechanical type system. Another approach to Jacobi metric can be found in \cite{LLM}.

\section{Comments on Jacobi metric and Hamilton-Jacobi equation}

Remark that condition $E(X)= (T+V\circ\tau_M)(X)=E_0\in\mathbb{R}$ is 
equivalent to say that the image of $X:M\to TM$ is contained in the hypersurface  of $TM$
$$\left\{v\in TM\mid  E(v)=T(v)+V(\tau_M(v))=\frac 12g(v,v)+V(\tau_M(v))=E_0\right\}\, .$$
This is a natural condition because the total energy $E=T+V$ is a constant of the motion for the dynamical Newtonian system. In fact, if  $X\in\mathfrak{X}(M)$ is a solution to the equation $\nabla_XX=-\textrm{grad\,} V$, then:
$$
\mathcal{L}_X(E(X))=\mathcal{L}_X\left(\frac 12g(X,X)+V\right)=g(\nabla_XX,X)+\mathcal{L}_XV\, ,
$$
and therefore, as 
$$
g(\nabla_XX,X)=g(-\textrm{grad\,} V,X)=-d V(X)=-\mathcal{L}_XV\, ,
$$
we obtain that $\mathcal{L}_X(E(X))=0$. 

Hence for every vector field $X\in\mathfrak{X}(M)$ solution to $\nabla_XX=-\textrm{grad\,} V$, we  have that the image of every integral curve of $X:M\to TM$ is contained in a constant  energy surface, that is, $E\circ 
\gamma:\R\to\mathbb{R}$ is a constant $E_0$. This constant is generically 
different for every integral curve.

The above Theorem can also  be  stated  for the curves solution to the mechanical type systems, that is: a curve satisfying $\nabla_{\dot\gamma}\dot\gamma=Z_F\circ\gamma= -(\textrm{grad\,} V)\circ\gamma$ is a pregeodesic of the metric  $\bar g=e^{2k} \exp(2\varphi)g$, with  $k\in\mathbb{R}$, and  in particular for the Jacobi metric determined by $k=0$.

For the general Newtonian case $\nabla_XX=Z_F$, where $F$ is an arbitrary  semi-basic 1-form on $TM$, the necessary and sufficient condition for 
$X$ to be a pregeodesic vector field is 
the existence of a function $f$ such that $\nabla_XX+g(X,X) X_F=f\, X$, 
 but  this situation  is not easy to deal with and  we have not any constant of the motion to be chosen instead of the energy.
Observe that if $E(X)=E_0$ the family of integral curves of $X$ is contained in the above hypersurface of $TM$.

The above results are related with  Hamilton-Jacobi equation.  In fact if we suppose that the vector field $X$ satisfies the equation $d\,( i_Xg)=0$, then condition $\nabla_XX=-\textrm{grad\,} V$ is equivalent to $d\,(E(X))=0$, which is the global expression of the classical Hamilton-Jacobi equation in the Lagrangian form. See \cite{CGMMMR06} for more details on this subject.

\section{Conclusions and future work}

We started with  the classical definition of Sundman transformation for autonomous systems of first-order differential equations and  its applications at the beginning of twentieth 
century and describe it in a geometric way as a conformal change in 
the corresponding. dynamical vector field. This geometric approach allows us to implement new applications 
and to understand why such transformations can be useful when additional geometric structures are present. The case of systems of second-order differential equations 
is not so easy  because the concept of velocity also changes. However,  those derivable from a variational principle as those
related to geodesic motions or Newtonian mechanical systems have  been studied  in  the framework of  Riemannian geometry. We can give a geometric answer to what is the relation between vector fields if we can relate their integral curves or we impose some particular properties to these 
integral curves. This is  stated and solve in Sections 5 and 6. 

As we relate the Sundman transformation with a conformal change, we use the same idea to obtain the Jacobi metric as a kind of Sundman transformation, in fact a conformal one of the metric of the manifold which is the configuration space of the mechanical system. This is made in Section 7, Theorem \ref{HJ-metric}, and the relation with Hamilton-Jacobi equation for Newtonian systems is commented in Section 8. This application is also 
related with the results obtained in Section 6.

There are several future lines of development, most of them related with the problems included in Section 3. On the other side, the geometric definition of Sundman transformation for general systems of second-order differential equation needs to be more carefully studied.  As the concept of velocity after a Sundman transformation is different, we have to consider  alternative tangent bundle structures.   This shows the convenience of   the study of the different structures associated with the tangent and cotangent bundles of a manifold, and this also affects to   the idea of linear and Hamiltonian systems respectively. We hope to present  some results in these lines in a forthcoming paper.

\section*{Acknowledgements}

\hspace{10mm}The authors 
acknowledge the financial support from the Spanish Ministerio de Ciencia, 
Innovaci\'on y Universidades projects
PGC2018-098265-B-C31 and PGC2018-098265-B-C33. 

\section*{Abbrevations}
Not applicable
\section*{Declarations}
\indent   

 {\bf Ethics approval and consent to participate:}

Not applicable.

\medskip

{\bf Consent for publication:}

Informed consent was obtained from all individual participants included in the study.

\medskip

{\bf Availability of data and material:}

Not applicable.

\medskip

{\bf Competing interests:}

The authors declare that they have no Competing interests.

\medskip

{\bf Funding:}

Financial support from the Spanish Ministerio de Ciencia, 
Innovaci\'on y Universidades,  projects
PGC2018-098265-B-C31 and PGC2018-098265-B-C33 (MINECO, Madrid) and DGA project E48/20R An\'alisis y F\'{\i}sica Matem\'atica.   

\medskip

{\bf Authors' contributions:}

The authors, JFC, EMF and MCML, equally contributed to the paper and have approved the submitted version

\medskip

\section*{Conflicts of interest}

The authors declare that they have no conflict of interest.

\end{document}